\documentclass{egpubl}
\usepackage{eurovis2026}

\SpecialIssuePaper         % uncomment for final version of Computer Graphics Forum, special issue
\CGFccby
\usepackage[T1]{fontenc}
\usepackage{dfadobe}  

\usepackage{cite}  % comment out for biblatex with backend=biber
\BibtexOrBiblatex
\electronicVersion
\PrintedOrElectronic
\ifpdf \usepackage[pdftex]{graphicx} \pdfcompresslevel=9
\else \usepackage[dvips]{graphicx} \fi

\usepackage{egweblnk}
\usepackage{tcolorbox}
\usepackage{ulem}
\usepackage{booktabs}
\usepackage{multirow}
\usepackage[table]{xcolor} 
\usepackage{array}
\usepackage{tabularx}
\usepackage{colortbl}
\usepackage[HTML]{xcolor}
\usepackage{graphicx}
\usepackage{tcolorbox}
\usepackage[utf8]{inputenc}
\usepackage{tikz}
\usepackage{adjustbox}
\usepackage{setspace}
\usepackage{hyperref}

\newcommand{\rev}[2]{{\color{black}{#2}}}

\newcommand{\rrev}[2]{{\color{black}{#2}}}

\title
{How Do LLMs See Charts? A Comparative Study on High-Level Visualization Comprehension in Humans and LLMs}
\author[H.Jeon et al]
{\parbox{\textwidth}{\centering Hyotaek Jeon$^{1}$\orcid{0009-0006-0030-6677}, Hyunwook Lee$^{1,2}$\orcid{0000-0002-5506-7347}, Minjeong Shin$^{1,3}$\orcid{0000-0001-6516-0433}, Tapendra Pandey$^4$\orcid{0009-0001-8641-6490}, Joohee Kim$^5$\orcid{0000-0002-0745-2339}, 
\\
Shinwook Seon$^1$\orcid{0009-0007-1942-4814}, Daeun Jeong$^1$\orcid{0009-0008-2258-9764}, Sungahn Ko\thanks{Corresponding Author}$^{1}$\orcid{0000-0002-7410-5652},
        and Ghulam Jilani Quadri$^{4}$\orcid{0000-0002-8054-5048} 
        }
        \\
{\parbox{\textwidth}{\centering $^1$Pohang University of Science and Technology (POSTECH), Republic of Korea\\
         $^2$Soongsil University, Republic of Korea\\
         $^3$  Australian National University, Australia \\
         $^4$University of Oklahoma, USA\\
         $^5$Ulsan National Institute of Science and Technology (UNIST), Republic of Korea\\
       }
}
}

\begin{document}

% uncomment for using teaser
% \teaser{
%  \includegraphics[width=0.9\linewidth]{eg_new}
%  \centering
%   \caption{New EG Logo}
% \label{fig:teaser}
%}

\maketitle
%-------------------------------------------------------------------------
\begin{abstract}
Designers often create visualizations to achieve specific high-level analytical or communication goals. These goals require people to extract complex and interconnected data patterns. Prior perceptual studies of visualization effectiveness have focused on low-level tasks, such as estimating statistical quantities, and have recently explored high-level comprehension of visualization. Despite the growing use of Large Language Models (LLMs) as visualization interpreters, how their interpretations relate to human understanding or what reasoning processes underlie their responses remains insufficiently understood. In this work, we explore LLMs' comprehension of visualization, examining the alignment between designers' communicative goals and what their audience sees. We have conducted a qualitative study to investigate the gap between human interpretative strategies and the reasoning pathways of LLMs across three types of visualizations, line graphs, bar graphs, and scatterplots, to identify the high-level patterns generated by LLMs using three prompt conditions. Our analysis results indicate that LLMs exhibit a consistent interpretative strategy that remains unchanged across prompt constraints. Furthermore, we observe two distinct approaches: humans naturally synthesize data into trend-centric narratives, whereas LLMs persist with a structural enumeration of comparisons and numerical ranges. Lastly, we see LLMs achieve visualization comprehension through mechanisms distinct from human intuition, pointing to critical challenges and new opportunities for visualization design. 
%-------------------------------------------------------------------------
%  ACM CCS 1998
%  (see https://www.acm.org/publications/computing-classification-system/1998)
% \begin{classification} % according to https://www.acm.org/publications/computing-classification-system/1998
% \CCScat{Computer Graphics}{I.3.3}{Picture/Image Generation}{Line and curve generation}
% \end{classification}
%-------------------------------------------------------------------------
%  ACM CCS 2012
%The tool at \url{http://dl.acm.org/ccs.cfm} can be used to generate
% CCS codes.
%Example:
\begin{CCSXML}
<ccs2012>
   <concept>
       <concept_id>10010147.10010178.10010179</concept_id>
       <concept_desc>Computing methodologies~Natural language processing</concept_desc>
       <concept_significance>500</concept_significance>
       </concept>
   <concept>
       <concept_id>10003120.10003145.10011768</concept_id>
       <concept_desc>Human-centered computing~Visualization theory, concepts and paradigms</concept_desc>
       <concept_significance>500</concept_significance>
       </concept>
   <concept>
       <concept_id>10003120.10003145.10011769</concept_id>
       <concept_desc>Human-centered computing~Empirical studies in visualization</concept_desc>
       <concept_significance>500</concept_significance>
       </concept>
 </ccs2012>
\end{CCSXML}

\ccsdesc[500]{Computing methodologies~Natural language processing}
\ccsdesc[500]{Human-centered computing~Visualization theory, concepts and paradigms}
\ccsdesc[500]{Human-centered computing~Empirical studies in visualization}

\printccsdesc   
\end{abstract}  
%-------------------------------------------------------------------------
% Sections

\section{Introduction}
Information visualizations help people extract meaningful analytical insights from data in the form of statistical quantities. 
For example, visualizations help people make sense of epidemiological data about COVID-19~\cite{kaul2020rapidly, borland2021enabling} or make predictions about natural disasters~\cite{cheong2016evaluating, preston2019uncertainty, millet2020hurricane}. 
For meaningful analytical insights, an effective visualization is needed that clearly communicates the designer’s intent to viewers, even when they are not directly cued to read off specific values or statistical quantities. 
The efficacy of such visualization depends on alignment between the designer's communicative intent and the viewer's intuitive understanding.
\rev{R1-1}{This concept, referred to as \textbf{High-Level Visualization Comprehension}, describes the holistic knowledge and salient patterns that viewers naturally extract from a chart without explicit cues or guidance~\cite{quadri2024you}.}
While prior work has characterized these interpretative pathways in humans~\cite{quadri2024you}, the reasoning mechanisms of LLMs in this context remain opaque.

With the advancement of Large Language Models (LLMs), we have observed their wide applications, including generating knowledge, comprehension, and insights for various chart types~\cite{zhao2025leva, zhao2025lightva, sultanum2023datatales}.
However, there is variance in the perceptual and cognitive abilities of LLMs compared to those of people~\cite{hong2025evaluation, wang2024aligned, stokes2025write}. 
Multimodal LLMs have enabled LLM agents to jointly process visual and textual information with unprecedented fluency~\cite {gpt4o,gemini}.
As these capabilities mature, the integration of LLMs into the visualization tasks is reshaping how people interact with data, extending their role beyond code generation~\cite{zhu2024automated, maddigan2023chat2vis} toward interpretation~\cite{podo2024vrecs}, critique~\cite{shin2025visualizationary, kim2025goodchatgptgivingadvice}, and analytical assistance~\cite{sah2024generatingnlq}.
It is crucial to determine whether LLMs possess intrinsic strategies for comprehending charts that mirror the human intuition and align with the designer's intent, or if they merely simulate understanding through high-speed, mechanical data retrieval and knowledge built on trained data~\cite{cheng2025understandingllm, hong2025evaluation, das2025chartsofthought}. 
As LLMs are increasingly integrated into the visualization pipeline, they are reshaping how users interact with data, acting not just as tools but as collaborative partners capable of reading and explaining visual information.
Without dissecting this decoding process and identifying whether LLMs genuinely see the same patterns as humans, leveraging them for collaborative analytics or automated design risks introducing subtle misalignments that could undermine the communicative goals of visualization~\cite{quadri2024you}.

In this work, we present a comparative study investigating the high-level visualization comprehension strategies of humans and LLMs. 
The objective is to holistically explore how LLMs and humans interpret graphs, organically and without any cues.
Using the question \textit{"Describe what do you see in the graph?"}, we have elicited descriptions from three MLLMs (e.g., GPT, Claude, and Gemini), across a balanced dataset of 60 visualizations established by Quadri et al.~\cite{quadri2024you}.
\rev{R4-1}{We design the experiment that varied visual stimuli across chart types, data types, and graph composition types to probe the models' adaptability to diverse visual structures.}
Furthermore, to disentangle the models' core reasoning capabilities from their tendency toward verbosity~\cite{llmasajudge}, we introduce varying prompt constraints, from unconstrained descriptions to single-sentence summaries. 
This approach allows us to observe how interpretative priorities shift under pressure and directly compare these LLM-generated descriptions with human high-level comprehension.
We identified three research questions: 
\begin{description}
    \item[{[RQ1]}]\textit{\rev{R4-1}{How do LLMs’ interpretation strategies vary across different prompt constraints and chart, data, and composition types?}}
    \item[{[RQ2]}]\textit{To what extent do human and LLM comprehension strategies diverge when comprehending visualizations?}
    \item[{[RQ3]}]\textit{How well do LLMs align with the designer’s intent compared to humans?}
\end{description}

To address these questions, we establish a qualitative and quantitative analysis process that maps responses to the hierarchical cognitive stages of Bloom's Taxonomy~\cite{bloom2,Burns20} and examines the statistical tasks~\cite{amar2005low, quadri2024you} (\autoref{framework}).
Our analysis reveals that LLMs exhibit a highly consistent interpretative strategy, prioritizing structural enumeration and data decoding, regardless of output length constraints.
Unlike humans, who naturally synthesize data into trend-centric narratives through pattern inference and intuitive sensemaking, LLMs persist with a mechanical strategy of listing comparisons and numerical ranges.
LLMs operate primarily through structural decoding and explicit value retrieval, focusing on reading individual coordinates rather than connecting them into a coherent insight.
Furthermore, while humans rely on visual scaffolding and can be hindered by diverse chart and composition types, LLMs demonstrate high visual faithfulness and a superior ability to align with the designer's intent. 
Our findings contribute a \rrev{[R1-2]}{process} for comparing visualization comprehension using statistical quantities, patterns, and Bloom’s Taxonomy; empirical evidence of the strategic divergence between human sensemaking and machine decoding; and insights into the potential of LLMs as objective evaluators in visualization comprehension pipelines. 
\section{Related Work}
\rev{R1-9}{In this section, we describe prior work on visualization tasks, human insight, and comprehension and LLMs for visualization.}

\subsection{Visualization Task, Insight, and Comprehension}
\rev{R1-2}{Modern design guidelines are grounded in graphical perception, the study of how people perceive specific information in visualizations~\cite{cleveland1984graphical} by performing visualization tasks~\cite{rensink2010perception, davis2024ranking}.
In contrast, high-level comprehension concerns the overall knowledge that viewers intuitively acquire about the data without explicit cues or guidance~\cite{quadri2024you}. 
Such comprehension reflects the salient statistics and patterns that organically emerge from a particular combination of data and design.
Viewers should be able to naturally derive insights and support the high-level analytical goals intended by the designer, since such insights are typically complex and go beyond low-level tasks ~\cite{amar2005low,quadri2021survey}. 
}

High-level comprehension often involves a bottom-up process, where visual attention is guided by the specific stimulus and its salient features, rather than solely by the viewer’s explicit goals or intentions~\cite{connor2004visual}. 
A survey by Yi et al.~\cite{yi2008understanding} identifies how people gain insights through visualization, informing the design of more effective visualizations and systems.
Further, they characterize four key steps: providing an overview, adjusting the view, detecting patterns, and matching a mental model.
Similarly, Webb’s Depth of Knowledge (DOK)~\cite{webb2002depth} framework categorizes learning into four levels of increasing cognitive complexity: recall and reproduction, application of knowledge and skills, strategic thinking, and extensive thinking. 
Another structured framework is Bloom's Taxonomy ~\cite{bloom2}, which is a hierarchical framework to categorize levels of comprehension and learning objectives. 
This taxonomy consists of knowledge, comprehension, application, analysis, synthesis, and evaluation.

\subsection{LLMs for Visualization Interpretation}

Recent progress in Large Language Models and integration of LLMs into visualization tasks is reshaping how people interact with data, extending their role beyond code generation to authoring, interpretation, critique, and analytical assistance. 
Recent studies focus on using LLMs to automate visualization design ~\cite{zhu2024automated}, for example, by generating visualization code, captioning, recommending design alternatives, often operating on natural language descriptions of data and tasks ~\cite{podo2024vrecs,ouyang2025nvagent,maddigan2023chat2vis, sultanum2023datatales, sah2024generatingnlq}. In visual analytics workflows, they assist with design feedback on existing charts ~\cite{shin2025visualizationary, kim2025goodchatgptgivingadvice}, help structure analytic tasks ~\cite{sah2024generatingnlq}, and support data exploration, insight generation, and storytelling \cite{zhao2025leva,zhao2025lightva,sultanum2023datatales}. 
Prior systems largely use LLMs as components within broader workflows, leaving their standalone role in visualization comprehension underexplored. Existing approaches emphasize syntactic correctness, not whether visualizations align with intended semantic or communicative goals—an issue heightened by LLM unreliability and the need for systematic behavioral understanding~\cite{cheng2025understandingllm}.

Knowing how well LLMs comprehend visualizations is critical for both LLM-assisted co-design ~\cite{ruan2025llmdesignstudy} and LLM-assisted analytics. 
This understanding can determine whether and how LLMs should be used as interpreters in visualizations. 
Recent work has begun to investigate the capabilities of LLMs for low-level visualization tasks, such as basic chart reading and performing visual analytics directly on chart representations, while assessing their visualization literacy and analytical skills ~\cite{bendeck2025gpt4vislit,das2025chartsofthought,hong2025evaluation}. 
However, there is still limited understanding of how LLMs respond to visualizations “in the wild” and how their interpretations align with the designer's intent, especially when charts are complex, embedded in realistic contexts, and designed to communicate specific messages to diverse audiences. 

\begin{figure}[t]
    \centering
    \includegraphics[width=0.83\columnwidth]{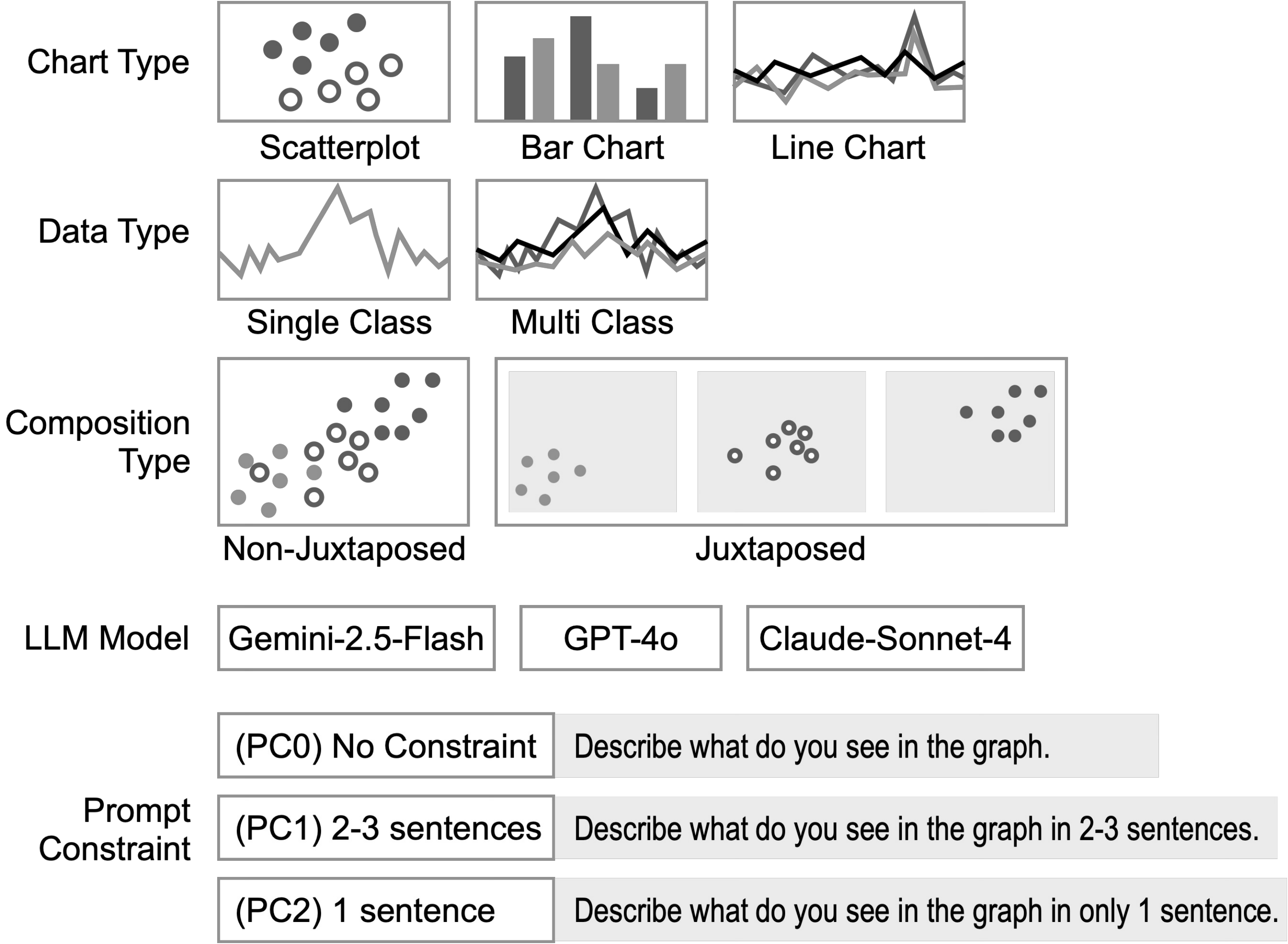}
    \caption{Study Design Dimensions: Chart Types(3), Data Types(2), Composition Types(2), LLMs(3), Prompt Constraints(3).}
    \vspace{-0.5cm}
    \label{fig:datatypefigure}
\end{figure}

\noindent \textbf{Human - LLM Comparison}
Recent work explored using LLMs as alternatives to human participants in empirical studies \cite{stokes2025write}, motivating comparisons of human and LLM chart interpretations to assess whether these models can serve as proxies for human viewers.
% -TP}
Focusing on low-level visualization tasks, current LLMs do not perform on par with human baselines. In particular, the LLMs fail to reach the levels of visualization literacy reported for the general public on VLAT-based assessments and rely heavily on prior knowledge and cued prompts for better performance ~\cite{hong2025evaluation, das2025chartsofthought, valentim2025plotthickens}. 
Similarly, Wang et al. ~\cite{wang2024aligned} study LLM behavior on bar charts to see how bar layout, chart context, and underlying data influence takeaways and to what extent these takeaways reflect the patterns and layout sensitivities observed in human interpretations. 
LLM takeaways are often not fully accurate and diverge from human comparison patterns, motivating alignment studies across more chart types. In exploring alignment between visualization‑specific takeaways (i.e., visualization affordances) of LLMs and humans, Stokes et al.~\cite{stokes2025write} found that GPT‑4o failed to match the human pattern of responses in unconstrained interpretations. In our work, we evaluate additional LLMs and focus on alignment with broader chart takeaways and insights, rather than limiting our evaluation to visualization affordances.

\section{Methodology}
In this work, we conduct a qualitative experiment to characterize the patterns and statistics LLMs comprehend in standard visualizations when they encounter a visualization without a guiding task. 
Using the prompt \textit{"Describe what do you see in the graph?"}, we elicit descriptions from three MLLMs (GPT, Claude, and Gemini), across a balanced dataset of 60 visualizations. 
Our methodology consists of two phases: (1) the generation of HLC descriptions by LLMs across varying chart types and constraints, and (2) a systematic comparison of these outputs with the human response dataset.

\subsection{Study Design}
Visualizations in real-world scenarios incorporate a wide array of designs, each potentially conveying distinct sets of statistical information. 
To design an ecologically valid experiment, we utilize the prior work's stimulus design~\cite{quadri2024you}. 
Next, we describe the stimuli and datasets.

\noindent \textbf{Stimuli:} 
\rev{R4-1}{We select three chart types, two data types, and composition types as independent variables as shown in \autoref{fig:datatypefigure}. }

\noindent \textit{Chart Type} includes bar charts, line charts, and scatterplots. 
Bar charts primarily support discrete comparisons between categories, while line charts facilitate identification of temporal trends and continuous changes. 
Scatterplots support analyses of correlation and clustering distributions. 
By examining these diverse types, we can assess how different visualization structures influence the interpretative strategies of both humans and LLMs.

\noindent \textit{Data Type} distinguishes between single-class data, which contains no categorical sub-groupings, and multi-class data, where information is decomposed into distinct categories via visual channels such as color or shape. 
Investigating this dimension enables us to analyze comprehension across both continuous and categorical encodings, thereby revealing how increased information density interacts with the visualization's communicative goals. 

\noindent \textit{Composition Type} contrasts non-juxtaposed designs with juxtaposed arrangements. 
Non-juxtaposed charts present data within a single coordinate system, such as superimposed lines or clustered bars, whereas juxtaposed designs utilize side-by-side or stacked small multiples. 
This distinction is methodologically significant because juxtaposed visualizations often impose a higher cognitive load, requiring the viewer to compare data and synthesize interpretations across multiple panels to understand the collective message. 

\noindent \textbf{Visualization and Natural Language Response Datasets} 
Based on the three primary design dimensions above, we use the dataset from \cite{quadri2024you}, which comprises 60 charts (five per unique combination of the three design dimensions). 
We choose this dataset for three reasons.
First, the chart designs are initially sourced from professional data journalism outlets to capture authentic visual structures used in practice. 
Second, to isolate visual decoding from semantic knowledge, the text labels and underlying data of the charts in the dataset are replaced with synthetic datasets, while the original visual structures of the charts are retained.
This ensures that the comprehension observed in both humans and LLMs is driven purely by the graphical representation, rather than by prior knowledge of specific real-world events or topics.
Third, the dataset includes natural-language responses to the prompt "Describe what do you see in the graphs" from 24 participants, comprising 18 females and 6 males, aged 18 to 56.
By combining these variables, we construct a comprehensive dataset comprising 60 unique stimuli (3 chart types × 2 data types × 2 juxtapositions × 5 charts) to rigorously evaluate the structural and narrative differences in humans' and LLMs' comprehension of visualizations.

\subsection{LLM Models and Prompt Strategies}
To facilitate a robust comparative analysis, we use three state-of-the-art models: GPT-4o~\cite{gpt4o}, Claude Sonnet 4~\cite{claude}, and Gemini 2.5 Flash~\cite{gemini}.
We choose these for their advanced vision-language capabilities, which we benchmark for comparing machine interpretation with human comprehension. 
To address model verbosity~\cite{llmasajudge}, we use prompts of varying lengths to assess models' capacity for information synthesis.
Our identified strategy leverages the finding that explicit constraints compel models to prioritize salient content and to adjust the granularity of information~\cite {controllablesummarization}.  
%: 1) No constraint (PC0), 2) 2--3 sentences (PC1), and 3) 1 sentence (PC2). 
We design three prompts with different length constraints, as shown in \autoref{fig:datatypefigure}.
The prompt for \textit{unconstrained condition (PC0)} presents models with the same open-ended question used in the human trials, thereby enabling natural generation strategies without artificial limitations. 
Subsequently, the prompt for the \textit{constrained condition (PC1)} limits responses to two or three sentences, forcing the models to filter visual information and prioritize essential elements. 
Finally, the prompt for the \textit{extreme synthesis condition (PC2)} restricts the output to a single sentence. 
This highly constrained setting is designed to uncover the model’s high-level perception, compelling it to abstract fine-grained details and articulate the single most significant pattern or message it identifies.

\definecolor{myDarkGray}{gray}{0.50} 
\definecolor{myLightBack}{gray}{0.96} 

\tcbset{
    stackbase/.style={
        colback=myLightBack,   
        colframe=myDarkGray,   
        coltitle=white,       
        fonttitle=\bfseries\sffamily,
        left=3mm, right=3mm, top=1.5mm, bottom=1.5mm,
        boxsep=0mm,
        toptitle=1.2mm, bottomtitle=1.2mm, 
        before skip=0pt,        
        after skip=0pt
    },
    firstbox/.style={
        stackbase,
        sharp corners=south,    
        rounded corners=north, 
        before skip=1em         
    },
    middlebox/.style={
        stackbase,
        sharp corners,          
        toprule=0pt             
    },
    lastbox/.style={
        stackbase,
        sharp corners=north,    
        rounded corners=south, 
        toprule=0pt,           
        after skip=1em        
    }
}

\subsection{Coding Procedures}
\noindent{\textbf{Statistical Task.}}
To ensure robust, scalable analysis of the generated descriptions, we employ a hybrid evaluation strategy that combines automated LLM-based coding with human verification.
For the initial coding of statistical tasks, we adopt a Human-AI collaborative workflow inspired by recent thematic coding analysis\cite{llmcoder1, llmcoder2}. 
Following these \rrev{[R1-2]}{process}, we use an LLM-as-a-Judge~\cite{llmasajudge} approach powered by GPT-5 to process the dataset while maintaining systematic consistency efficiently. 
The model is prompted to classify each sentence of the descriptions into statistical tasks, with detailed definitions provided in \autoref{framework}.
To assess the reliability and potential bias, two of the authors independently coded 10\% of the data without access to the LLM’s results. 
The LLM’s coding is then compared to the two human coders’ results. 
Finally, we observe agreement of 97.3\% (Fleiss’ Kappa = 0.922) between the human and LLM coding results. 

\noindent{\textbf{Bloom's Taxonomy.}}
\rrev{[R1-1]}{Parallel to the statistical task extraction, we characterize the cognitive states exhibited in the descriptions by mapping each response to a specific level within Bloom’s Taxonomy.
Whenever the analytical actions and cognitive behaviors are observed (e.g., 'Describe a trend' for \textit{Analysis} or 'Explain the topic' for \textit{Comprehension}), the response is assigned to the corresponding level.
To ensure the reliability of our mapping, six independent coders first annotate a 10\% random subset of the data following partial-overlap methodologies~\cite{Krippendorff04,artstein-poesio-2008-survey}. 
After resolving discrepancies through iterative discussions and majority voting, we obtain an inter-coder reliability of Krippendorff's $\alpha = 0.87$, allowing the remaining data to be evenly divided and coded individually.
}
\vspace{-10pt}

\noindent{\textbf{Visual Faithfulness.}}
To empirically determine whether the models' reasoning results in factually accurate descriptions or leads to hallucinations unsupported by the visual evidence, three human coders evaluated the visual faithfulness~\cite{faithfulness1,faithfulness2} of the responses.
For the faithfulness assessment, coders verify the extent to which each description is grounded in the source visualization and classify responses into three distinct levels, \textit{Perfect Faithful, Partial Error}, or \textit{Unfaithful / Hallucination} (detailed in \autoref{sec:appendixA}). 
The coding sessions yield an agreement level of 84.63\%.
To account for heavily skewed data distributions, we utilized Gwet’s AC1~\cite{gwet2008computing} alongside Fleiss’ Kappa. 
The assessment indicated high reliability (Gwet’s AC1 = 0.82) and confirmed the overall factual reliability of the chosen models (86.7\%--Fully Faithful, 12.7\%--Partial Errors, 0.6\%--Major Error), ensuring that subsequent analyses rely on factually accurate descriptions.
 
\noindent{\textbf{Designer's Intent.}}
While visual faithfulness ensures factual accuracy, we further evaluated alignment with the designer's intent using a four-level scale.
\textit{Complete Match} indicates that the model captured both specific statistical patterns and overarching insights, aligning fully with the design objective. 
\textit{General Match} refers to responses that convey the correct core message and general trends but lack precise statistical details or granular values. 
\textit{Partial Match} is assigned when the model identifies the general subject matter (e.g., variables or axes) but misses the primary trends or relationships intended by the design. 
Finally, \textit{No Match} indicates a complete failure to comprehend the chart's purpose, leading to interpretations that do not align with the stated objective.
The three human coders achieve high inter-rater reliability (Gwet’s AC1 = 0.85, $\kappa$ = 0.74), with any disagreements resolved through discussion.

\begin{figure*}[t]
    \centering
    \includegraphics[width=0.9\textwidth]{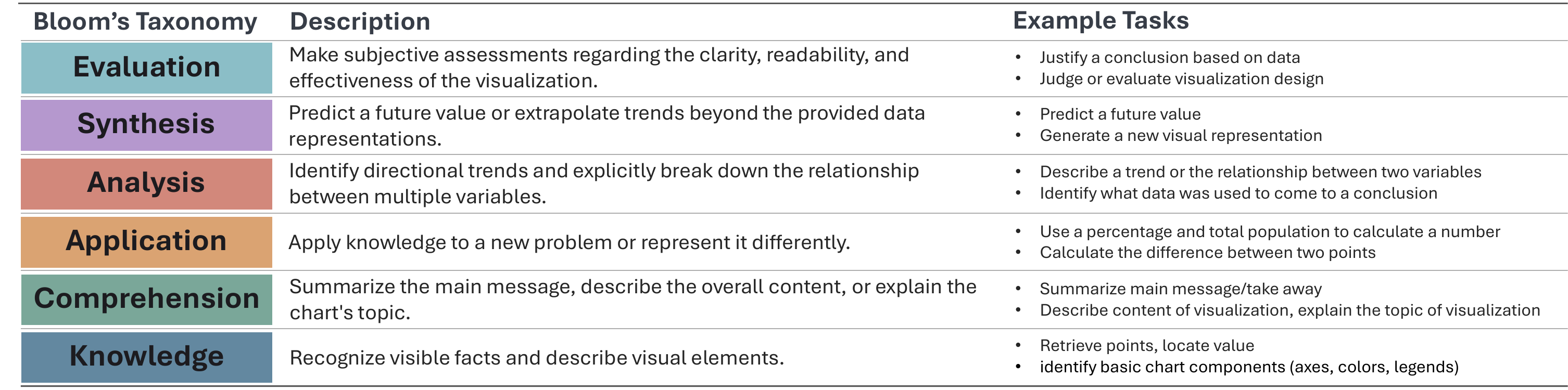}
    \caption{\rev{R2-3, R3-8, R4-4}{Bloom’s Taxonomy for visualization comprehension with descriptions and example tasks~\cite{Burns20}.}}
    \label{fig:bloomstaxonomy}
    \vspace{-0.4cm}
\end{figure*}

%\vspace{-10pt}
\subsection{Semantic Similarity Analysis of LLMs' Descriptions}
\label{sec_semantic_sim}
To empirically validate whether the models maintain consistent interpretations despite the varying prompt length constraints, we employ semantic textual similarity analysis~\cite{textualsimilarity}.
This metric measures semantic equivalence, quantifying how closely core meanings align regardless of verbosity.
To analyze LLM descriptions, we generate high-dimensional vector representations of responses using OpenAI's \textit{text-embedding-3-large} model~\cite{neelakantan2022text}, which translates texts into numerical vectors to quantify their semantic similarity.
We then calculate the cosine similarity between the embedding vectors of the descriptions (e.g., PC0 vs. PC2).
The similarity score between two vectors $\mathbf{v}_1$ and $\mathbf{v}_2$ is defined as:
\begin{equation}
    Similarity(\mathbf{v}_1, \mathbf{v}_2) = \frac{\mathbf{v}_1 \cdot \mathbf{v}_2}{\|\mathbf{v}_1\| \|\mathbf{v}_2\|} = \frac{\sum_{i=1}^{n} v_{1,i} v_{2,i}}{\sqrt{\sum_{i=1}^{n} v_{1,i}^2} \sqrt{\sum_{i=1}^{n} v_{2,i}^2}}
\end{equation}
where $\mathbf{v}_1$ and $\mathbf{v}_2$ represent the embedding vectors of the descriptions generated under different constraints. 
This metric quantifies the stability of the model's comprehension, revealing whether the core semantic meaning remains consistent or drifts as the response length is constrained. 
We average these scores across all samples to derive overall semantic robustness for each model and chart type.

\subsection{Comparative Evaluation of Comprehension}
 
To systematically assess the divergences in high-level comprehension, we constructed a multi-dimensional evaluation \rrev{[R1-2]}{process} that integrates quantitative analysis with qualitative interpretive assessments. 
We conducted a comparative analysis along three strategic dimensions: statistical quantities analysis, cognitive complexity via Bloom's taxonomy, and the designer's intent alignment.

\uline{\rev{R1-8}{Statistical quantities analysis}} characterizes the interpretative focus of humans and LLMs, revealing differences in how they process visual information.
Utilizing a closed taxonomy, we analyzed the frequency and distribution of tasks (e.g., \textit{Trend}, \textit{Comparison}) to determine whether the descriptions are dominated by element-wise data extraction or holistic pattern synthesis. 
\rrev{[R1-1]}{\uline{Bloom's taxonomy} is examined to assess interpretative sophistication. 
By mapping the responses to the Bloom's Taxonomy (as detailed in \autoref{framework}), we quantify the cognitive complexity of the interpretations.
Specifically, we contrast the proportion of lower-order cognitive levels (e.g., \textit{Knowledge}) against higher-order levels (e.g., \textit{Analysis}) to determine if LLMs remain at the level of superficial data decoding or achieve insight generation.}
\uline{Designer's intent alignment} determines the pragmatic validity of the interpretations. 
We benchmark models in discerning the chart's communicative purpose against human performance, distinguishing between merely technically accurate descriptions and those effectively articulating the intent.
\section{\rrev{[R1-2]}{Analytic Procedures for Visualization Comprehension}}\label{framework}

\noindent \textbf{Statistical Tasks}
\rev{REV}{We code the descriptions using a closed taxonomy of nine low-level statistical tasks adapted from prior work~\cite{quadri2024you, amar2005low}.
This taxonomy includes: \textit{Correlation (Cor), Comparison (Comp), Trend (Tr), Cluster (Clu), Compute Derived Value (CDV), Determine Range (DR), Extremum (Xtrm), Characterize Distribution (CD)}, and \textit{Anomaly (Anom)}.
Since a single response often incorporates a variety of these operations, we identify multiple tasks per response, preserving their chronological sequence to trace the logical flow of the interpretation.}

\noindent \textbf{\rrev{[R1-1]}{Bloom's Taxonomy}} 
%%%%%%%%%%%%% new version
\rrev{}{has been used in visualization research to distinguish different levels of understanding.
Burns et al.~\cite{Burns20} showed that visualization evaluation should account for multiple levels of understanding, and their subsequent work further demonstrated that these levels may vary depending on visual representation~\cite{Burns22}. 
In visualization education and literacy research, related work has extended this perspective to onboarding, learning goals, and the design and evaluation of literacy-oriented modules~\cite{Stoiber2023,Byrd20197d,Peng2022}.
More recently, it has been used to assess how LLMs recommend teaching visualization techniques~\cite{Joshi2024}. 
These studies suggest that visualization comprehension is not only in terms of correctness or value extraction, but also in terms of the level at which understanding is expressed in responses.
This is particularly relevant in our setting, where we compare open-ended chart descriptions and therefore need to distinguish whether a response remains at fact retrieval, summarizes overall meaning, or interprets relationships and structure.}

\rrev{}{In our procedure, Bloom’s Taxonomy is used to characterize the level of understanding expressed in each response. 
To this end, we map each response to a specific level within Bloom’s Taxonomy~\cite{Burns20,Burns22}, and provide the descriptions and illustrative examples for each level in \autoref{fig:bloomstaxonomy}. 
Because a single response may express multiple levels of understanding, we allow multiple Bloom categories to co-occur. 
This analysis is conducted separately from the statistical task coding, while statistical task coding captures the types of analytical operations expressed in a response, Bloom coding characterizes the level at which the response is articulated.
The goal of this coding is not to infer latent cognitive processes directly, but to describe the level of understanding expressed in the response in a systematic and comparable way.}

\noindent \textbf{Thematic Coding }
\rev{R1-4}{The lead author of this work codes the mapping reasons by developing low-level themes in a codebook.
Then, two other authors review the themes and revise them together to see if there is any disagreement with the proposed themes. 
Once all themes are established, the three authors independently re-code the data using the themes and subsequently aggregate their results.
}

\section{Theme 1: LLMs Maintain Consistent Reasoning Mechanisms Regardless of Output Constraints and \rev{[R4-1]}{Types of Charts, Data, and Composition in Visualization}}
\label{sec-theme1}

We systematically investigate the interpretative strategies of LLMs by quantifying information volume and performed statistical tasks across varying \rev{R4-1}{prompt constraints (see \autoref{fig:datatypefigure} and types of charts, data, and composition of visualization}).
Furthermore, we assess semantic robustness and consistency of the generated responses to verify if the underlying comprehension remains stable.

\begin{figure}[t]
    \centering
    \includegraphics[width=\columnwidth]{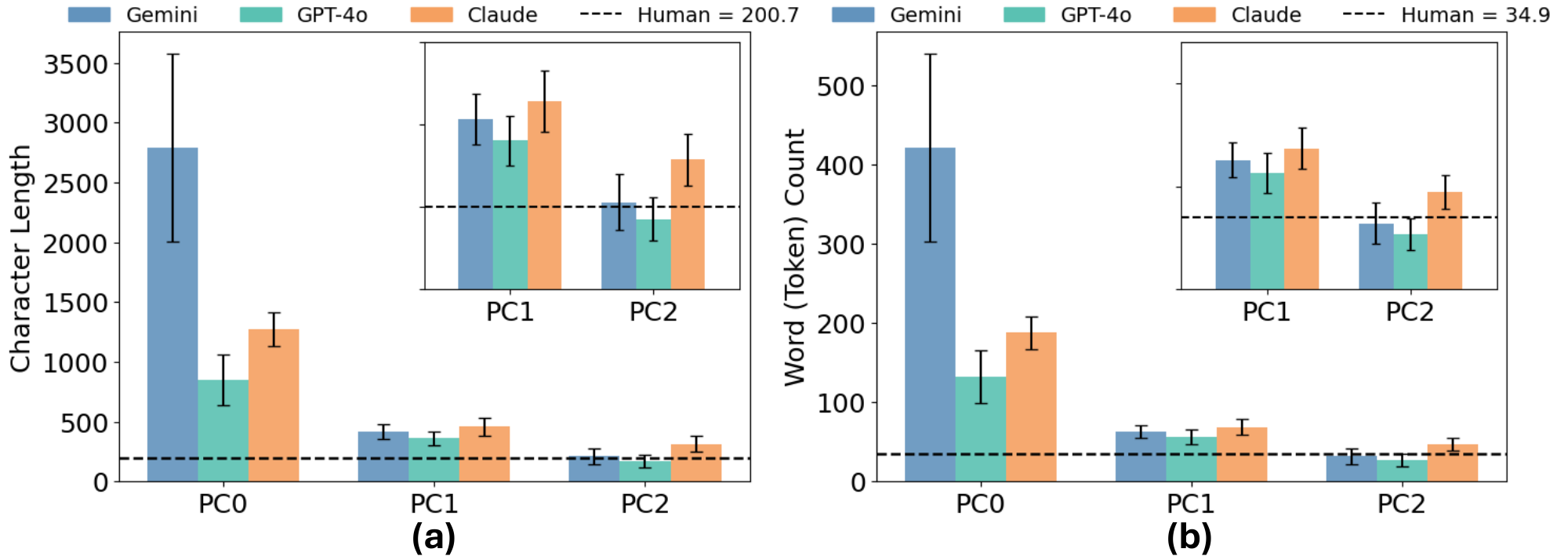}
    \caption{Character Length (a) and Word (Token) Count (b) of description generated for three prompt constraints (PC0, PC1, PC2).} 
    \label{fig:tokencount}
    \vspace{-0.4cm}
\end{figure}
\subsection{LLMs Prioritize Data over Narratives}

LLMs consistently employ an exhaustive decoding strategy to maximize information density.
We first examine the output characteristics of humans and LLMs with two metrics: \textit{response lengths} and \textit{statistical tasks} counts using the nine task taxonomy defined in \autoref{framework}.
The most immediate distinction appears in the description-text volume. 
As shown in \autoref{fig:tokencount} (a), under the unconstrained condition (PC0), LLMs generate significantly longer descriptions in terms of both the word and character counts, compared to humans.
This means LLMs tend to produce elaborative descriptions that reflect a difference in processed information density.
Notably, when the prompt is constrained to a single sentence (PC2), the length of the LLM-generated responses converges to a level very similar to that of natural human responses.
This indicates that while humans instinctively compress information, LLMs require explicit constraints to produce information volumes comparable to humans.

\renewcommand{\arraystretch}{0.85}
\begin{table}[t]
    \centering
    \caption{Average statistical task counts described by LLMs across visualization, data, and composition types, and prompt constraints.}
    % LLM 평균값이다 ~ 
    \label{tab:statisticaltask_avgcount}
    \setlength{\tabcolsep}{15pt}
    \resizebox{\columnwidth}{!}{%
    \begin{tabular}{lccc}
        \toprule
        \textbf{Category} & \textbf{PC0} & \textbf{PC1} & \textbf{PC2} \\
        \midrule
        \multicolumn{4}{c}{\textit{\textbf{Visualization Type}}} \\
        \midrule
        Line Chart      & 5.00 & \textbf{3.57} & \textbf{2.90} \\
        Bar Chart       & 4.35 & 2.95 & 2.53 \\
        Scatterplot     & \textbf{5.15} & 3.23 & 2.45 \\
        \midrule
        \multicolumn{4}{c}{\textit{\textbf{Data Type \& Composition Type}}} \\
        \midrule
        Single Class    & 4.62 & 3.28 & 2.57 \\
        Multi Class     & 4.93 & \textbf{3.47} & \textbf{2.71}  \\
        Single Class Juxtaposed   & \textbf{5.15} & 3.26 & 2.64\\
        Multi Class Juxtaposed   & 4.62 & 2.97 & 2.58 \\
        \midrule
        \multicolumn{4}{c}{\textit{\textbf{Model Type}}} \\
        \midrule
        Gemini-2.5-Flash   & \textbf{5.43}  &  2.98  & 2.23  \\
        GPT-4o   & 4.05 &  3.18  & 2.47  \\
        Claude-Sonnet-4   & 5.02 &  \textbf{3.58} & \textbf{3.18} \\
        \bottomrule
    \end{tabular}
    }
    \vspace{-0.3cm}
\end{table}

Prompt constraints influence the high-level comprehension description of charts with tasks, as shown in \autoref{tab:statisticaltask_avgcount}.
LLMs operating without constraints (PC0) generate descriptions containing an average between 4.05 (GPT4o) and 5.43 (Gemini) statistical tasks.
Furthermore, even under the single-sentence constraint (PC2), which yields response lengths most similar to humans, LLMs still perform significantly more statistical tasks, averaging 2.23 (Gemini) and 2.47 (GPT-4o), compared to the human average of 1.53. 
This demonstrates that, unlike humans, who filter information before articulation, LLMs employ an exhaustive decoding strategy, attempting to translate as many visual features as possible into a description.
This also indicates that, unlike humans, who translate visual patterns into a high-level narrative, LLMs lack semantic prioritization and instead focus on decoding specific values.

\begin{table}[t]
\centering
\caption{\rev{R3-2}{Average count of statistical tasks by different types of charts, data and composition  (SC: Single Class, MC: Multi Class, J: Juxtaposed) under three prompt constraints.}}
\label{tab:model_task_count}
\resizebox{\columnwidth}{!}{%
\begin{tabular}{c|ccc|ccc|ccc} % 첫 번째 컬럼 l -> c 변경
\toprule
% 첫 번째 행: 모델명 (3칸씩 병합)
\multirow{2}{*}{\textbf{Chart Type}} & \multicolumn{3}{c|}{\textbf{GPT-4o}} & \multicolumn{3}{c|}{\textbf{Gemini-2.5-Flash}} & \multicolumn{3}{c}{\textbf{Claude-Sonnet-4}} \\ \cmidrule{2-10}
% 두 번째 행: 세부 조건 (PC0, PC1, PC2)
 & PC0 & PC1 & PC2 & PC0 & PC1 & PC2 & PC0 & PC1 & PC2 \\ \midrule
Bar Chart & 3.55 & 2.90 & 2.45 & 4.95 & 2.70 & 2.10 & 4.55 & 3.25 & 3.05 \\ \midrule
Line Chart & 4.15 & \textbf{3.45} & \textbf{2.70} & 5.45 & \textbf{3.15} & \textbf{2.45} & \textbf{5.40} & \textbf{4.10} & \textbf{3.55} \\ \midrule
Scatterplot & \textbf{4.45} & 3.20 & 2.25 & \textbf{5.90} & 3.10 & 2.15 & 5.10 & 3.40 & 2.95 \\ 
\midrule
\midrule
SC  & 3.66 & 3.06 & 2.26 & 5.06 & 3.13 & \textbf{2.40} & 5.13 & 3.66 & 3.06 \\ \midrule
MC   & 4.06 & 3.20 & 2.26 & \textbf{5.66} & \textbf{3.40} & 2.33 & 5.06 & \textbf{3.80} & \textbf{3.53} \\ \midrule
SC-J & \textbf{4.53} & \textbf{3.40} & \textbf{2.73} & \textbf{5.66} & 2.86 & 2.26 & \textbf{5.26} & 3.53 & 2.93 \\ \midrule
MC-J & 3.93 & 3.06 & 2.60 & 5.33 & 2.53 & 1.93 & 4.60 & 3.33 & 3.20 \\ \bottomrule
\end{tabular}%
}
\end{table}

\subsection{LLMs Compress Descriptions According to Narrative Complexity \rev{R4-1}{and Types of Charts, Data, and Composition}}

The LLM's tendency toward enumeration exhibits an interesting reversal depending on the \rev{R4-1}{visualization type} and prompt constraints. 
\autoref{tab:statisticaltask_avgcount} shows that under the unconstrained condition (PC0), LLMs perform the most tasks for scatterplots (5.15).
However, under conditions where summarization is forced (PC1, PC2), the task count for line charts remains higher than for scatterplots. 
\rev{R3-2}{\autoref{tab:model_task_count} presents a more granular inter-model comparison. While models generally show high task counts for scatterplots under PC0, they consistently produce higher task counts for line charts than for scatterplots under PC1 and PC2.
}
Regarding data types, single-class juxtaposed charts trigger the most tasks under PC0, whereas multi-class charts record the higher task counts under PC1 and PC2 on average.

\begin{figure}[t]
    \centering
    \includegraphics[width=\columnwidth]{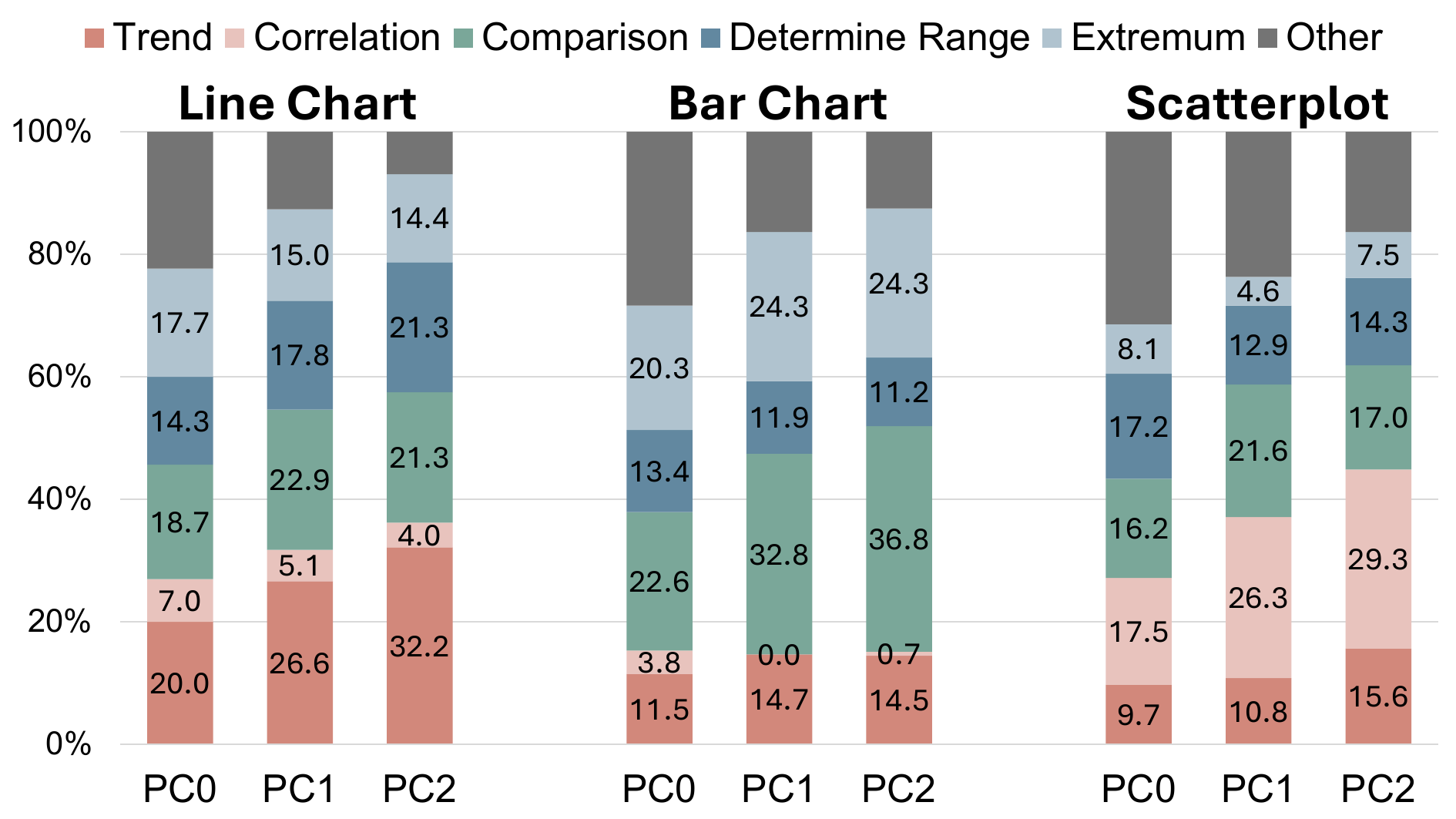}
    \caption{\rev{R1-6}{The distribution of statistical tasks across chart types and prompt constraints.}} 
    \label{fig:summary_strategy}
    \vspace{-0.5cm}
\end{figure}

\autoref{fig:summary_strategy} illustrates how LLMs \rev{R3-3}{realign}
in response to tightening length constraints (PC0 $\rightarrow$ PC1 $\rightarrow$ PC2). 
We observe that the prioritization of statistical tasks varies significantly across chart types.
For line charts, the most prominent shift occurs in the extraction of global patterns. 
As the constraint intensifies with PC2, the proportion of \textit{Trend} tasks increases notably. 
This suggests that when space is limited, LLMs prioritize the overall temporal directionality as the most critical information for line charts.
In contrast, \rev{R3-3}{bar charts maintain stable \textit{Trend} proportions, with LLMs allocating more space to \textit{Comparison} tasks instead.}
This reflects the discrete nature of bar charts, where comparing magnitudes between categories is often more semantic than identifying a continuous trend.
\rev{R3-3}{LLMs show the most dramatic shift with Scatterplots.
The relative proportion of \textit{Correlation} tasks expands significantly in PC2, indicating that the relationship between variables is prioritized as the defining feature of scatterplots.
}

We find that when unconstrained, LLMs tend to enumerate in response to the sheer number of visual objects (e.g., the numerous points in a scatterplot). 
\rev{R3-3}{
However, under strict length constraints, LLMs demonstrate a capacity to compress these responses into a single high-level descriptor, such as "positive correlation."
This behavior likely reflects a reaction to the inherent affordances of the visualization types themselves.
Since scatterplots are explicitly engineered to elicit correlational insights, the LLM’s shift represents a predictable response to these strong visual signals.
While line charts or bar charts posses their own inherent affordances, these visual signals are arguably less singular compared to the correlation in a scatterplot.
For instance, describing a trajectory in a line chart entails a higher narrative complexity, requiring a minimum set of components--start point, peak, end point, volatility--that cannot be easily omitted.
}
In other words, LLMs enumerate when visual information volume is high (scatterplots), but struggle to summarize when narrative difficulty is high (line charts), resulting in relatively more tasks being mentioned. 
Humans, conversely, maintain a relatively consistent cognitive complexity across chart types by increasing abstraction as complexity increases.

\subsection{LLMs Describe Charts with Semantic Robustness and Task Consistency Across Prompt Constraints}

\noindent \textbf{Model Robustness} assesses the semantic stability of the model's comprehension as output constraints are relaxed.
We measure this by calculating the semantic textual similarity (\autoref{sec_semantic_sim}) between responses generated under different constraints.
Our analysis reveals high stability across all models. 
The similarity scores for the PC2 $\rightarrow$ PC1 transition are consistently high, ranging from 0.875 (Gemini) to 0.893 (GPT-4o). 
Even in the extreme transition from a single sentence to an unconstrained response (PC2 $\rightarrow$ PC0), the models maintain robust similarity scores between 0.826 and 0.837.
Although the unconstrained response condition shows the lowest similarity, as expected given the significant differences in length and detail, the scores remain well above 0.8, indicating that the core semantic message does not drift significantly but expanded.

\noindent \textbf{Consistency in Statistical Quantities} investigates information retention by measuring the percentage of core tasks preserved when generating a more detailed response. 
Results demonstrate exceptional consistency (\autoref{Table/taskconsistancy}, Appendix). 
All models retain over 86\% of core tasks when expanding from PC2 to PC1. 
Notably, Gemini exhibits the highest consistency, retaining 97.62\% of tasks in PC2 $\rightarrow$ PC0 and 94.98\% in PC1 $\rightarrow$ PC0. Across visual forms, consistency remains high. 
For PC1 $\rightarrow$ PC0, scatterplots achieve the highest average (95.10\%), suggesting their explicit features are decoded reliably. 
Bar and line charts also exhibit robust consistency (90.44\% and 90.36\%). 
This uniformity across chart types indicates that LLMs prioritize and retain the specific statistical tasks essential for each chart’s interpretation throughout the elaboration process.

\section{Theme 2: Humans Derive Narratives While LLMs Decode Structures From  Charts}
% Humans Leverage Visual Scaffolding While LLMs Treat Visualizations as Raw Data.??
\label{sec-theme2}

To deeply understand how LLMs comprehend visual information in charts, we compare their generated descriptions to those of humans using Bloom's Taxonomy. 
We also investigate the specific types of statistical tasks prioritized by each model under the single-sentence condition (PC2) to isolate strategic differences from verbosity, where response lengths are most similar to those of humans.
Our results reveal distinct differences in how human and machine agents comprehend charts.
Next, we present our key findings.

\begin{figure}[]
    \centering
    \includegraphics[width=0.9\columnwidth]{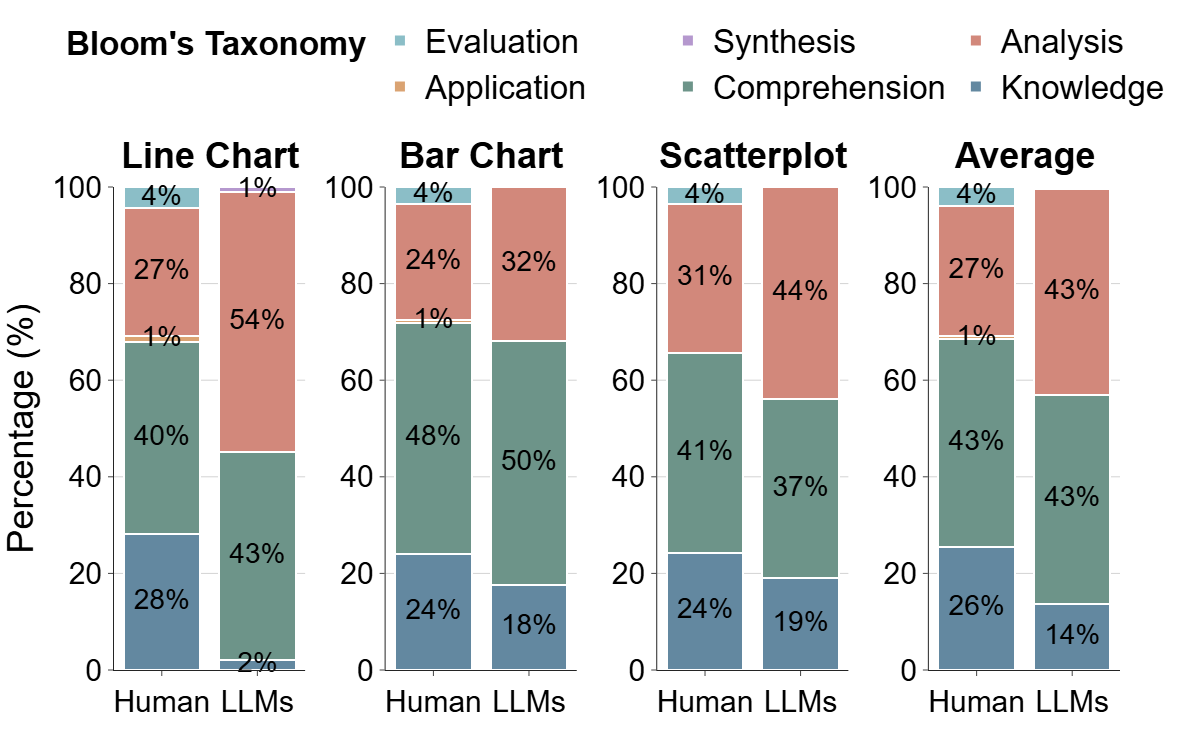}
    \caption{\rev{R2-3, R3-8, R4-4, R4-11}{Distribution of Bloom's Taxonomy cognitive categories across different chart types for humans and LLMs.
    %It shows the final cognitive categories reached via task sequences across Bloom’s Taxonomy by humans and LLMs side by side, given the same stimuli and tasks.
    }}
    %\vspace{-0.4cm}
    %Pairs of Stacked bars from left to right show comparison in a)bar charts, b)line charts, c)scatterplot and d) overall average, respectively. 
    \label{fig:bloomsratio}
    \vspace{-0.5cm}
\end{figure}

\subsection{\rev{R2-3, R3-8, R4-4}{Comparative Distribution of Cognitive Strategies}}
\label{sec:6.1}

\rev{R2-3, R3-8, R4-4, R4-11}{
Comparing the overall distribution of the categories reveals distinct differences in how humans and LLMs allocate cognitive resources during visualization comprehension. 
As shown in \autoref{fig:bloomsratio}, on average, both groups utilize the \textit{Comprehension} category at similar rates to grasp the overall meaning of a chart. 
However, detailed cognitive patterns exhibit clear divergence. 
Humans frequently utilize the \textit{Knowledge} category (26\%) to extract individual data points and occasionally perform \textit{Evaluation} to assess the validity of the visualization. 
Conversely, LLMs exhibit a lower proportion of \textit{Knowledge} and a complete absence of \textit{Evaluation}, instead showing a pronounced concentration in \textit{Analysis} (43\% vs. humans' 27\%).

These divergent characteristics are consistently observed across various chart types. 
In line charts, humans maintain a proportional balance between \textit{Knowledge} and \textit{Analysis}, whereas LLMs allocate a mere 2\% to \textit{Knowledge} with \textit{Analysis} reaching 54\%. 
This pattern holds for bar charts and scatterplots, where human Knowledge utilization remains steady, and LLMs consistently demonstrate a strong skew toward \textit{Analysis}. 
Consequently, the distribution indicates that LLM outputs are predominantly concentrated on analytical concepts, whereas human responses exhibit a broader distribution that includes foundational data extraction.

Next, we analyze how multiple cognitive categories are mapped to a single response (multi-label patterns), because it provides a deeper understanding of the differences in reasoning progression between humans and LLMs. 
For example, a response--"A bar graph which shows the population of Afghanistan per calendar year. The bars are blue." can be mapped to \textit{Comprehension} for identifying the overall topic and \textit{Knowledge} for extracting specific visual details. 
Our analysis results indicate that human responses exhibit a diverse and evenly distributed range of combinations, with no single pattern dominating. 
The most frequent patterns include single \textit{Comprehension} (21.2\%), \textit{Knowledge + Comprehension} (20.8\%), \textit{Comprehension + Analysis} (19.1\%), and \textit{Knowledge + Comprehension + Analysis} (11.8\%). 
These results suggest that rather than relying on a rigid template, humans adapt their cognitive strategies, employing diverse combinations of summary, data extraction, and analysis depending on the chart's specific visual context.

In contrast, LLM response patterns underline a distinct tendency to converge on a few specific combinations.
While the exploration and context-building patterns that appeared in diverse proportions in humans are diminished, the \textit{Comprehension + Analysis} (34.4\%) and \textit{Knowledge + Comprehension + Analysis} (22.2\%) patterns account for more than half of the LLM responses. 
The single \textit{Analysis} (20.6\%) pattern is also more than twice as frequent as in humans (8.3\%).
This consistent focus on \textit{Analysis} indicates a structurally uniform approach. 
Unlike humans, who utilize diverse reasoning strategies, LLMs generally approach chart interpretation as a standardized data decoding process.
}

\definecolor{Burgundy}{HTML}{9C4C46} %4
\definecolor{DeepBlue}{HTML}{A16F3C} %3
\definecolor{RichTeal}{HTML}{4E8079} %2
\definecolor{DeepGreen}{HTML}{55708C} %1

\begin{table}[t]
\centering
\resizebox{\columnwidth}{!}{%
\begin{tabular}{c c c c}
\toprule
\multicolumn{2}{c}{\textbf{Statistical Task (Human)}} &
\multicolumn{2}{c}{\textbf{Statistical Task (LLM)}} \\
\midrule
\textbf{Task} & \textbf{Count} &
\textbf{Task} & \textbf{Count} \\
\midrule

Comparison & \textbf{61} &
Comparison & \textbf{118} \\

Trend & \textbf{48} &
Trend & \textbf{101} \\

Correlation & \textbf{47} &
Determine Range & \textbf{75} \\

Compute Derived Value & 32 &
Extremum & 73 \\

Extremum & 29 &
Correlation & 51 \\

Anomaly & 10 &
Characterize Distribution & 24 \\

Characterize Distribution & 8 &
Cluster & 15 \\

Cluster & 4 &
Anomaly & 12 \\

Determine Range & 4 &
Compute Derived Value & 4 \\

\bottomrule
\end{tabular}%
}
\caption{Comparison of statistical tasks frequency in LLM and human description on the chart's comprehension.  %\ghulam{We can sort by bringing the top 3 above}
}
\label{tab:taskcount_rank}
\vspace{-0.5cm}
\end{table}

% \begin{table}[t]
% \centering
% \resizebox{\columnwidth}{!}{%
% \begin{tabular}{c c c c}
% \toprule
% \multicolumn{2}{c}{\textbf{Statistical Task (Human)}} &
% \multicolumn{2}{c}{\textbf{Statistical Task (LLM)}} \\
% \midrule
% \textbf{Task} & \textbf{Count} &
% \textbf{Task} & \textbf{Count} \\
% \midrule

% Trend \Analyze{\textbf{L4}} & 91 &
% Comparison \Understand{\textbf{L2}} & 118 \\

% Extremum \Remember{\textbf{L1}} & 62 &
% Trend \Analyze{\textbf{L4}} & 101 \\

% Comparison \Understand{\textbf{L2}} & 61 &
% Determine Range \Remember{\textbf{L1}} & 75 \\

% Correlation \Analyze{\textbf{L4}} & 52 &
% Extremum \Remember{\textbf{L1}} & 73 \\

% Compute Derived Value \Apply{\textbf{L3}} & 18 &
% Correlation \Analyze{\textbf{L4}} & 51 \\

% Characterize Distribution \Apply{\textbf{L3}} & 9 &
% Characterize Distribution \Apply{\textbf{L3}} & 24 \\

% Cluster \Analyze{\textbf{L4}} & 8 &
% Cluster \Analyze{\textbf{L4}} & 15 \\

% Determine Range \Remember{\textbf{L1}} & 4 &
% Anomaly \Analyze{\textbf{L4}} & 12 \\

% Anomaly \Analyze{\textbf{L4}} & 3 &
% Compute Derived Value \Apply{\textbf{L3}} & 4 \\

% \bottomrule
% \end{tabular}%
% }
% \caption{Comparison of statistical tasks mentioned by humans and LLMs. \textbf{Bloom's taxonomy level; L1 (Determine Range), L2(Comparison), L3 (Compute Derived Value), L4 (Trend)} \textcolor{teal}{Please check the supplementary section as PC0 to PC2 have been added.} %\ghulam{We can sort by bringing the top 3 above}
% }
% \label{tab:taskcount_rank}
% \end{table}

% \vspace{-10pt}
\subsection{\rev{REV}{Sequential Transition Pathways in Chart Comprehension}}
\label{sec:6.2}

In this section, we examine how humans and LLMs differ in the statistical tasks they prioritize during chart comprehension. 
\autoref{tab:taskcount_rank} compares the frequency of statistical tasks in LLM-generated description and human response.
Both groups begin with \textit{Comparison} and \textit{Trend}, reflecting a shared entry point into the visualization. 
% After these initial tasks, 
After these tasks, humans more frequently incorporate \textit{Correlation} and \textit{Compute Derived Value}, extending their interpretation toward relational and derived patterns. 
LLMs instead move toward \rev{REV}{enumerative data retrieval} after \textit{Trend}, relying heavily on \textit{Determine Range} and \textit{Extremum}, which involve direct retrieval of axis and point-wise values.
This difference reflects distinct comprehension strategies in which humans prioritize narrative synthesis and pattern-level inference while LLMs default to structural enumeration anchored in explicit coordinate extraction. 
The gaps in \textit{Determine Range} further highlight that LLMs are disproportionately engaged in \rev{REV}{isolated data extraction} that humans rarely perform.

\begin{figure}[t]
    \centering
    \includegraphics[width=0.8\columnwidth]{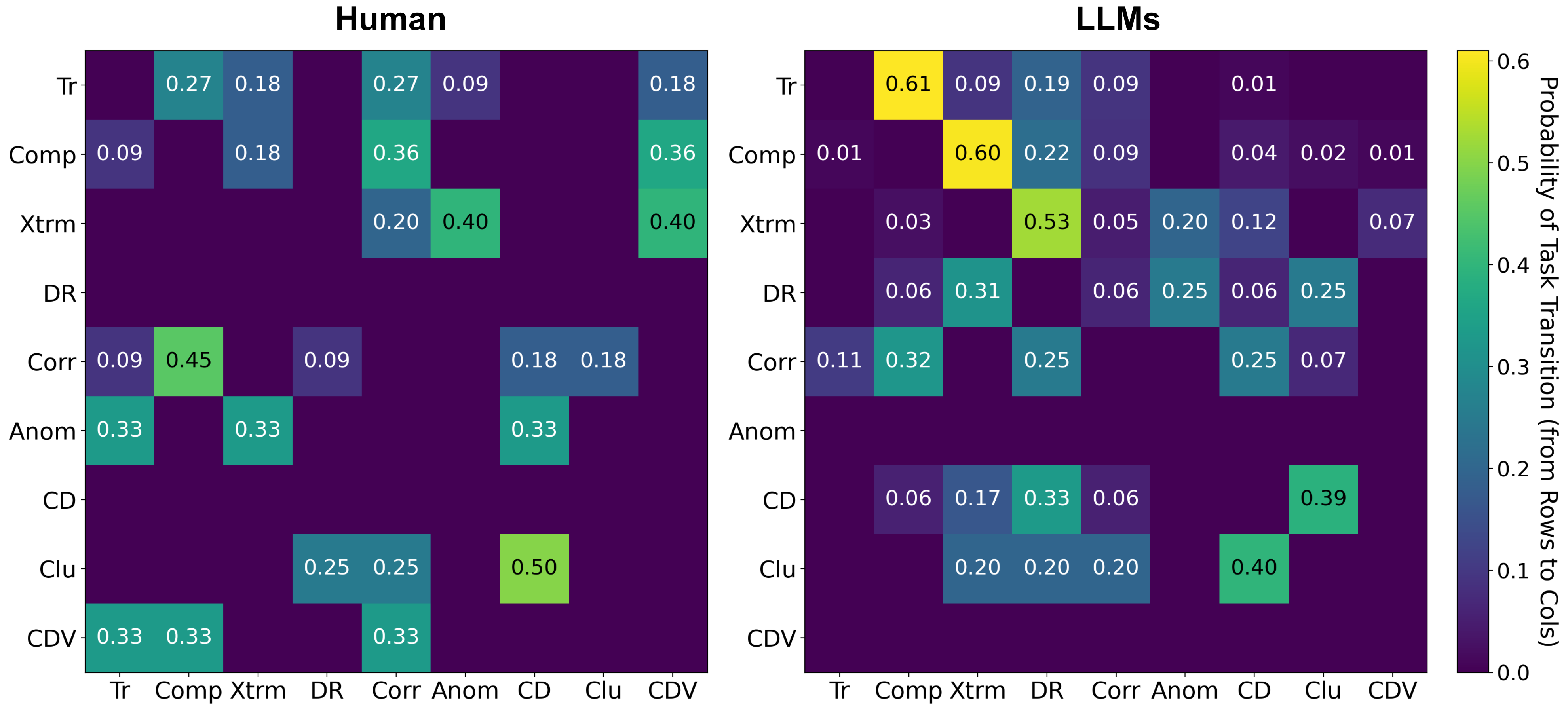}
    \caption{\rev{R1-6}{Task transition probability matrices for human and LLM descriptions. Cell values indicate the likelihood of transitioning from the row task to the column task.}
    }
    \label{fig:task_transition}
    \vspace{-0.5cm}
\end{figure}

These differences become more evident when examining the sequence in which humans and LLMs perform tasks. 
\autoref{fig:task_transition} illustrates the probability of task transitions; each cell in the heatmap represents the probability of transitioning from one task in the row to another task in the column. 
To compute these probabilities, we have extracted the chronological sequence of tasks from each description containing at least two tasks and calculated the frequency of adjacent task pairs.
For instance, in the case of LLMs, there is a 61\% probability of transitioning from a \textit{Trend} task to a \textit{Comparison} task, and a 60\% probability of transitioning from a \textit{Comparison} task to a \textit{Extremum} task. 
This implies that while \textit{Trend} does not necessarily appear as the initial task, once it occurs, it is highly likely to be followed by \textit{Comparison}, which in turn is frequently followed by \textit{Extremum}.
% C3G WWW Diagram Penn unversity 
We hypothesize that the observed sequence (\textit{Trend} $\rightarrow$ \textit{Comparison} $\rightarrow$ \textit{Extremum} $\rightarrow$ \textit{Determine Range}) stems from the structural conventions of chart captions and alternative text predominantly found in the LLM's pre-training data. 
Standard accessibility guidelines~\cite{WAIGuideline} and high-quality captioning datasets typically advocate for a top-down descriptive structure. 
This pattern also aligns with Shneiderman's Visual Information Seeking Mantra~\cite{Shneiderman96}: ``Overview first, details-on-demand.''

In contrast, human descriptions do not exhibit such a single dominant sequential chain.
While LLMs demonstrate a distinct preference for transitioning from \textit{Trend} to \textit{Comparison}, humans exhibit a more balanced distribution, showing comparable probabilities for proceeding from \textit{Trend} to either \textit{Comparison} or \textit{Correlation}.
Additionally, we observe notable transitions such as \textit{Correlation} $\rightarrow$ \textit{Comparison} (45\%) and \textit{Cluster} $\rightarrow$ \textit{Characterize Distribution} (50\%).
\rev{[REV]}{Unlike LLMs' mechanical reliance on basic data enumeration, humans connect broader patterns (e.g., correlations or clusters) with relational comparisons to build a cohesive interpretation.}

%\vspace{-10pt}
\subsection{The Role of Visual Composition and Scaffolding in Shaping Interpretative Focus}
\begin{figure}[t]
    \centering
    \includegraphics[width=0.8\columnwidth]{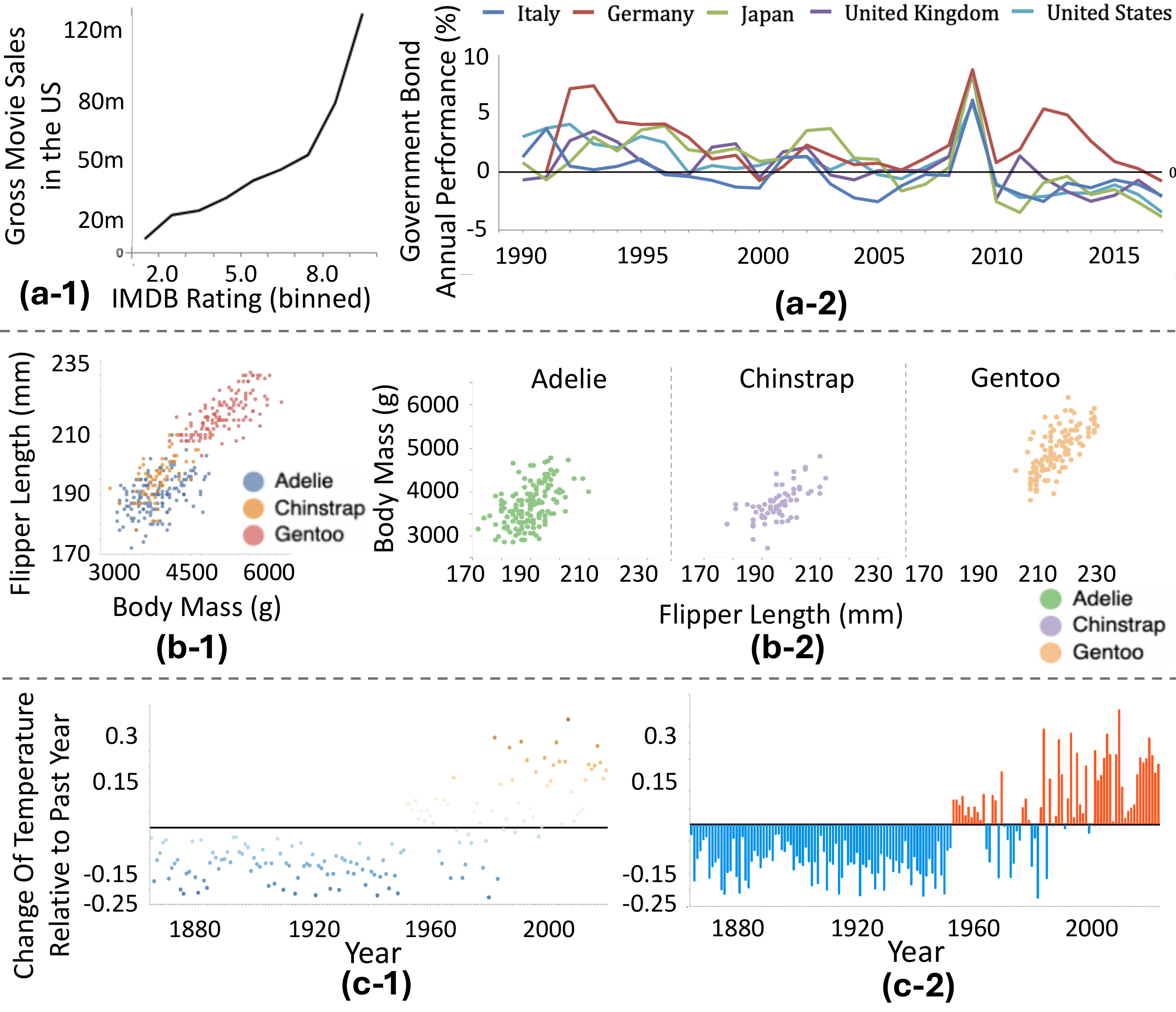}
    \caption{\rev{[R4-2, R1-6]}{Representative visualization examples used in our study: (a) line charts with (a-1) single-class and (a-2) multi-class data; (b) (b-1) non-juxtaposed and (b-2) juxtaposed layouts with the same data; and (c-1) a scatterplot and (c-2) a bar chart for chart type comparison with the same data.}}        \label{fig:differencegraph}
    \vspace{-0.5cm}
\end{figure}

In our analysis, we find that LLMs treat charts as raw data streams, generating identical descriptions regardless of layout changes.
To further investigate this aspect, we compare responses to pairs of visualizations with identical underlying data but differing composition (e.g., superimposed vs. juxtaposed) and encoding (e.g., bar chart vs. scatterplot).
First, we observe that humans actively identify and describe instances of visual overlap and structural indistinctness within a chart.
\autoref{fig:differencegraph} (b-1) shows an example of a scatterplot where multiple data points from different categories are heavily overlapped, causing visual uncertainty.

With the chart, humans tend to focus more on visual uncertainty during chart comprehension (e.g., "(Chinstrap is) intermingled with Adelie around the 3000-4000g range").
When analyzing charts in juxtaposed layout (\autoref{fig:differencegraph} (b-2)), their focus shifts to structural clarity. 
Humans often preface their analysis with remarks regarding the layout and visual clarity, explicitly stating that it is \textit{"easier to read, shows the correlation... per species in individual graphs."}
This structural change allows them to identify clear distinctions and to perform visualization tasks, stating that \textit{"Gentoo was the highest for both categories."}
In contrast, LLMs exhibit no such comprehension shift. 
For both the overlapped chart and its juxtaposed version, they provide nearly the same description, such as \textit{"Gentoo penguins being the largest… while Adelie and Chinstrap overlap."}
While visual separation serves as a critical aid to human comprehension, LLMs treat the chart merely as data, showing no qualitative change in their comprehension strategy regardless of the layout.

We find a similar divergence in comprehension that humans anchor their descriptions in the visual components within charts and LLMs rely on data values.
For example, when analyzing temperature data in a scatterplot (\autoref{fig:differencegraph} (c-1)), humans anchor their descriptions in the visual attributes such as color and encoding, noting \textit{"as the dots get higher up... they become a warmer color."}
Similarly for the bar chart (\autoref{fig:differencegraph} (c-2)), humans focus on visual properties, stating \textit{"blue bars and red bars"}. 
Conversely, LLMs generate descriptions that are semantically indistinguishable across chart types, characterizing the \textit{"clear shift from predominantly cooler years... to increasingly frequent and larger warmer years"} regardless of the visual form. 
This confirms again that humans interpret and aggregate the visualization components for comprehension, while LLMs focus on decoding the underlying the data points.

\section{Theme 3: LLM's Designer's Intent Alignment Does Not Require Human-Like Comprehension Strategies}
\label{sec-theme3}
In this section, we compare the models' descriptions are aligned with designer intent, measuring their capacity to identify the chart's communicative goals relative to human performance.
We assess the alignment with which LLMs extract the primary communicative goal against a ground truth dataset of 60 designer objectives. 
For the assessment, three of the authors classify each response into four categories~\cite{quadri2024you}: \textit{Complete Match, General Match, Partial Match}, or \textit{No Match}.
These objectives are constructed as compound analytical tasks, ranging from simple documentation (e.g., "\textit{To document the stock price of Google...}") to complex comparative inferences (e.g., "\textit{To show correlation/relation/association between a car odometer readings and their ages ...}").
A full summary of comprehension match counts is presented in Appendix \autoref{tab:comprehension_match}.

Our analysis indicates that LLMs' descriptions are better aligned with designer intent, achieving a high \textit{Complete Match} in 71.1\% (128/180) of cases, whereas human participants achieve this level in only 40.6\% (117/288). 
This disparity is distinctly reflected in the rates of intent misalignment. 
Humans exhibit a high \textit{No Match} rate of 27.1\% (78/288), indicating frequent difficulties in identifying the statistical quantities. 
In contrast, LLMs show no instances of \textit{No Match} (0\%), yet we observe a smaller \textit{Partial Match} cases (9.4\%, 17/180). 
LLMs generate descriptions with high-level comprehension that consistently are aligned with the designer's intent, but the descriptions often diverge from the analytical purpose.

Examining a \textit{Partial Match} case clarifies this divergence, with a multi-class juxtaposed scatterplot showing BMI and age across four regions (\autoref{fig_major_error} (c) in Appendix), 
The designer's intent is \textit{"to show and compare the age and BMI... and demonstrate the distribution... among 4 regions."} 
The LLM provides a mechanically precise description, stating the chart \textit{"displays four scatter plots showing client's BMI against client's age... across the southeast regions."} 
While factually accurate, this response does not fully align because it merely lists the visual components without executing the intended comparison or characterizing the distributions.
Conversely, humans struggle with the visual density of the same chart, focusing on surface attributes like \textit{"Lots of pink and blue dots"} (\textit{Partial Match}) or expressing heavy cognitive overload, such as \textit{"Too much, the overlap of color over the graph... too overstimulating"} (\textit{No Match}). 
This contrast suggests that human failure cases tend to arise from perceptual overload, whereas LLM's \textit{Partial Match} cases result from structurally shallow analysis.
These failure cases seem different, but they all lead humans and LLMs to incomplete comprehension.
Consequently, LLMs tend to treat visualizations as structured data schemas that they can readily decode, while humans rely more heavily on visual perception and thus remain more vulnerable to the cognitive load imposed by dense or complex designs.

\section{Discussion}
We have examined the high-level patterns that Large Language Models naturally produce when interpreting visualizations without an assigned task. 
Our work provides an initial step toward characterizing the unprompted patterns that arise from a graph’s design and their alignment with humans' high-level comprehension. 
Future work should build on these insights to develop heuristics that better support LLM-based visualization comprehension.

\noindent \textbf{Reflection on Prior Findings}
Our study results show that LLM-driven chart comprehension diverges from the human comprehension patterns characterized by Quadri et al.~\cite{quadri2024you}. 
While humans use visual scaffolding, LLMs exhibit a persistent preference for the structural enumeration of data, regardless of visual composition. 
This behavior suggests that the layout insensitivity reported by Wang et al.~\cite{wang2024aligned} stems from a broader processing difference in how models decode visual information.
Furthermore, while structured prompting strategies~\cite{das2025chartsofthought} have proven effective for enhancing low-level data extraction, our results show over-reliance on such scaffolding may inadvertently bypass the visually grounded reasoning strategies central to human-like sensemaking.

\rev{R2-4}{
Prior work has leveraged Bloom’s taxonomy to characterize levels of graph comprehension \cite{Burns20}. 
Building on this foundation, our work extends this approach by applying the six levels of the taxonomy to systematically analyze and characterize LLM-generated responses to high-level visualization comprehension tasks. 
In addition, prior work \cite{quadri2024you} demonstrates that human chart comprehension is strongly shaped by contextual information and representational expressiveness, reinforcing earlier findings \cite{Burns22} that contextual cues and visualization design play a role in shaping interpretation and understanding, which we observed in comparison between LLM and human responses.}

\noindent \textbf{Applying the Findings with Caveats}
An implication of our findings is the potential risk of using LLMs as proxies for human evaluators in the design of visualizations. 
Our study reveals a superior alignment with \textit{designer's intent} for LLMs (71.1\% \textit{Complete Match}) compared to humans (40.6 \% \textit{Complete Match}), suggesting they are effective at objectively validating whether a chart technically encodes the intended information. 
However, this numerical precision obscures a fundamental misalignment with human perceptual experience. 
As evidenced by our dense scatterplot analysis, LLMs effectively decode data that human participants have difficulty to process. 
This indicates that while LLMs may act as an effective evaluation tool for data validation, they are poorly aligned with the human perspective of \textbf{effective visualization}. 
As such, relying solely on LLMs to evaluate visualization effectiveness risks prioritizing numerical accuracy over human cognitive process, potentially leading to designs that human audience may not prefer.

\rev{[R2-6]}{\noindent \textbf{About Designer's Intent}
In this work, we define designers’ intent through an analytical lens, focusing specifically on communicative goals related to structured data extraction, following the framing proposed by Quadri et al.~\cite{quadri2024you}. 
Under this definition, we provide a structured model by narrowing our focus to the analytical objectives of visualization. 
However, we acknowledge that this represents only one layer of designers’ broader communicative intents in visualization, which may also include emphasis, framing, or audience impact—dimensions where humans may appear to fail analytically but still succeed communicatively.
}

\noindent \textbf{LLM Hallucination in High-Level Comprehension}
We have only one error across the entire dataset with a single-class bar chart characterized by 12 categories, as shown in \autoref{fig_major_error} (a) in Appendix. 
GPT-4o describes the chart as \textit{"Feedgrains have the highest harvest at around 26,000 tons, while melons and cotton \& wool have the lowest"}.
However, this description does not align with the visual data, as the actual highest bar corresponds to `Bakery Products' and the lowest to `Fungi'. 
Notably, the chart uses 12 indistinguishable colors, which does not follow a general guideline on the maximum number of colors~\cite{Ware2012}.
The fact that the model falters inaccurately where human perception would also struggle suggests a shared vulnerability to inappropriately designed charts.

\section{Limitations and Future Work}
We discuss limitations and future work extended from the study. 
\rev{R3-1}{We use Gemini-2.5-Flash, a state-of-the-art model optimized for efficiency. 
While it demonstrates satisfactory performance on the visualization comprehension task, we expect that more recent models may yield different results. }
\rev{R3-4}{Moreover, the prompts constraining the response length may inadvertently bias the models toward their training alignment for detailed captioning rather than reasoning. }
\rev{R3-5, R4-1, R4-3, R2-1}{Alternative prompting strategies (e.g., Chain-of-Thought, few-shot prompting) with different visualizations could help us investigate whether the observed mechanical patterns stem from an inherent reasoning limitation of the models or a response bias triggered by prompt configurations.

\rrev{[R2]}{
Visual complexity has been defined in multiple ways in prior work and may be approached either as a perceptual property of visual stimuli or through computational proxies~\cite{Donderi2006,Machado2015}.
In this study, chart type, data multiplicity, and composition are not treated as direct measures of visual complexity, but as design dimensions that may contribute to perceived complexity and interpretive difficulty.
Structural properties of a visualization may also introduce visual uncertainty, particularly in dense displays~\cite{Dasgupa2012}. 
Future work should therefore incorporate explicit and validated measures of visual complexity and examine how such measures interact with high-level visualization comprehension.}
While our \rrev{}{Bloom’s taxonomy and statistical task based coding process} serves as a useful descriptive heuristic, it may overlook important dimensions in human responses, such as uncertainty, affective reasoning, and critique of the visualization itself. 
Investigating these dimensions further could help us better understand the key components of high-level visualization comprehension.
 
}

\rev{R2-2}{We use collaborative coding with an LLM for the analysis. 
While we consistently observe qualitative contrasts between human narrative responses and LLM enumerations, the effectiveness of this collaborative coding approach may not be fully generalizable to other studies, as the results produced by this method are largely descriptive. 
To validate the coding results, we repeat human coding (Section 3.3). 
We think future in-depth analyses could help more directly demonstrate the significance of the coding results. }

\section{Conclusion}

In this work, we present a comparative study investigating the high-level visualization comprehension strategies in LLMs and humans. 
We examine chart comprehensions of LLMs with different chart types, data complexities, and prompt constraints.
Our analysis reveals a distinct divergence in comprehension approaches where humans prioritize to construct narratives while LLMs consistently rely on structural enumeration and explicit data extraction.
This tendency indicates that LLMs process charts primarily as structured data rather than as visual stories, yet they demonstrate a superior ability to align with designer intent by identifying communicative goals even in complex visualizations where human viewers often struggle. 
These findings suggest that they offer a distinct and valuable role as objective agents for validating design objectives and ensuring informational clarity in visualization workflows.

\noindent{\textbf{Acknowledgments}}\\
We are grateful to Cindy Xiong Bearfield (Georgia Tech) for insightful discussions and constructive feedback. 
This work was supported by the Institute of Information \& Communications Technology Planning \& Evaluation (IITP) grant (No. RS-2019-II191906, Artificial Intelligence Graduate School Program (POSTECH), the National Research Foundation of Korea (NRF) grant (No. RS-2024-00456247; No. RS-2023-00218913) funded by the Korea government (MSIT), and Korea Health Technology R\&D Project grant through KHIDI (HI22C0646) funded by the Ministry of Health \& Welfare.
%\newpage
%-------------------------------------------------------------------------
% bibtex
\bibliographystyle{eg-alpha-doi} 
\bibliography{egbibsample}       

% biblatex with biber
% \printbibliography                
%-------------------------------------------------------------------------
\setstretch{0.9715}
\appendix
\clearpage
\setcounter{section}{0}

\renewcommand{\arraystretch}{1} % Reset table height to default value (i.e., one)

\begin{center}
    {\Large \textbf{Appendices}}
\end{center}
\vspace{1em}

\section{\textbf{Visual Faithfulness Assessment}}~\label{sec:appendixA}
% 아래는 한문장으로 줄여도?
To evaluate the reliability of the generated descriptions, we assess visual faithfulness, defined as the extent to which the textual description is grounded in and supported by the source visualization~\cite{faithfulness1, faithfulness2}. 
For this, three of the authors have manually coded each response and categorized them into three distinct levels of faithfulness, \textit{Perfect Faithful, Partial Error}, or \textit{Unfaithful / Hallucination}. 
\textit{Fully Faithful} descriptions are defined as those that are factually accurate and entirely supported by the visual evidence without ambiguity. 
\textit{Partially Faithful} descriptions capture the general visual trend but suffer from imprecision, containing minor numerical inaccuracies or vague quantifiers. 
Finally, \textit{Unfaithful/Hallucination} refers to descriptions that fail to align with the visual data, including extrinsic fabrications (unsupported claims) or intrinsic contradictions.
Overall results are in Table~\ref{supp_VisualFaithfulness}.
%\sako{HT: here, need to add definition and examples of perfect faithful, partial error, unfaithful.}
%We manually coded each response and categorized them into three distinct levels of faithfulness, \textit{Perfect Faithful, Partial Error}, or \textit{Unfaithful / Hallucination}.

The evaluation results reveal that LLMs generally align generated descriptions with visual evidence accurately. 
In the result, we see that 156, out of 180 responses (86.7\%) are categorized as \textit{Fully Faithful}, reflecting the visual features and data trends while strictly adhering to the depicted information.
However we see that actual ratios slightly vary across chart types. 
The match score with line charts shows the highest fidelity (91.7\%, \textit{Fully Faithful}). 
The few partial errors observed are primarily due to difficulties in precise value extraction where multiple trends overlapped, or minor inaccuracies in interpreting shared axes across juxtaposed sub-charts. 
Scatterplots similarly show high precision (86.7\%), though occasional inaccuracies occur when isolating individual outliers within dense clusters or when the model slightly overestimate the strength of correlations in loosely distributed data.

\begin{figure*}[t]
    \centering
    \includegraphics[width=\textwidth]{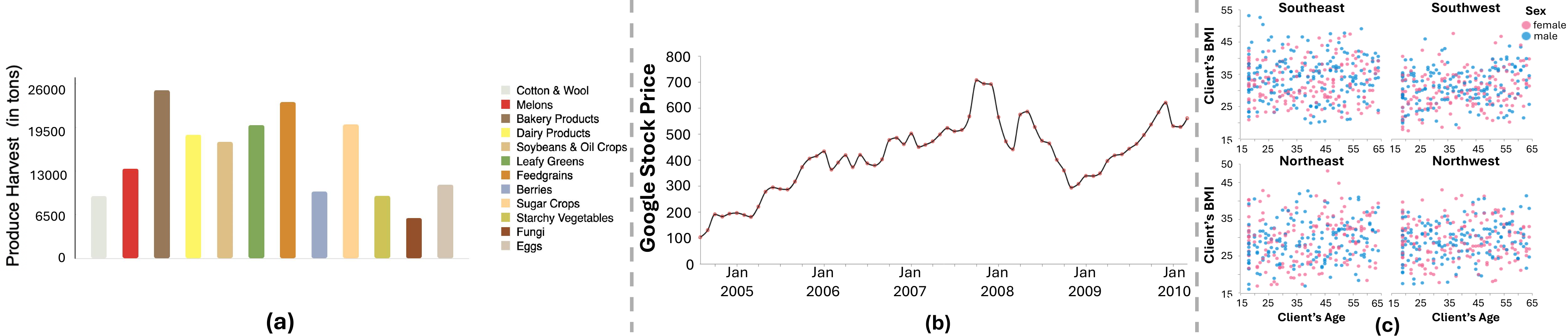}
    \caption{Examples of visualizations used in our study: (a) Single-class bar chart utilizing 12 distinct categories. (b) Single-class line chart displaying Google stock price. (c) Multi-class juxtaposed scatterplot visualizing between client's age and BMI across four regions.}
    \label{fig_major_error}
\end{figure*}

Scores with bar charts present distinct challenges in visually crowded environments. 
While the models consistently capture the overarching narratives and global trends, they exhibit minor deviations in fine-grained comparisons (16.7\% \textit{Partial Error}). 
These instances typically occur in multi-class or juxtaposed settings, where the models effectively summarize the distribution but occasionally simplify the ranking of adjacent bars with nearly identical heights.
We observed a single major error in a 12-category bar chart, as shown in Figure~\ref{fig_major_error} (a). 
The model incorrectly identified "Feedgrains" as the highest value and "Melons" and "Cotton \& Wool" as the lowest, whereas the actual maximum and minimum were "Bakery Products" and "Fungi," respectively.
This hallucination appears to be caused by the chart's use of 12 indistinguishable colors, which created visual ambiguity in identifying specific categories.

\section{\textbf{Detailed Analysis of Statistical Task Frequency}}
This section provides a granular breakdown of the statistical tasks performed by LLMs. While the main manuscript identifies a general tendency toward structural enumeration, this supplementary analysis dissects how specific design dimensions (Chart Type and Data Complexity) modulate this behavior.

\noindent \textbf{Task Frequency Variation Across Chart Types}
Table~\ref{tab:model_task_count} presents the average count of statistical tasks across different visualization types. 
The results highlight distinct interpretation strategies for each chart form, with Line Charts exhibiting a persistence to summarization.
Line Charts consistently elicit a high volume of statistical tasks across all models and constraints. 
Even under the strictest single-sentence constraint (PC2), models like Claude maintain an average of 3.55 tasks. 
This can be attributed to the inherent complexity of line chart; unlike other charts, describing a trend effectively often requires utilizing a large subset of the 9 available statistical tasks simultaneously, such as identifying the start and end points (\textit{Extremum}), defining the overall direction (\textit{Trend}), and noting fluctuations or ranges (\textit{Determine Range}). 
Consequently, the models struggle to compress this multi-faceted information into a low-task summary.

In contrast, Scatterplots show the highest volatility. They trigger the highest task counts in the unconstrained condition (e.g., 5.90 for Gemini) due to the enumeration of individual points but drop drastically in the constrained condition (e.g., 2.15). 
This suggests that while scatterplots appear visually dense, their semantic message can be easily compressed into a single task (e.g., \textit{Correlation}), unlike the irreducible temporal components of a line chart.
Bar Charts generally demonstrate a lower volume of statistical tasks compared to other types. 
This lower frequency reflects the discrete nature of the data, where interpretation is typically anchored in straightforward tasks such as identifying the maximum/minimum (Extremum) or performing simple pairwise Comparisons. 
Unlike the sequential integration required for line charts, the semantic structure of bar charts allows models to capture the core message through fewer, more distinct statistical operations.

\noindent \textbf{Data Complexity and Composition}
Following the chart type analysis, we examine how data density and layout composition influence model (Table~\ref{tab:model_task_count}). 
Our analysis reveals that increasing complexity does not linearly increase task enumeration; instead, it triggers a strategic shift towards abstraction.

\begin{table*}[t]
\centering
\caption{Visual faithfulness assessment across chart types. 
Of the 180 evaluated responses, 86.7\% were classified as Fully Faithful. 
Line charts exhibited the highest faithfulness, scatterplots remained highly reliable, 
and bar charts showed more partial errors due to fine-grained comparison difficulty in dense charts. 
One major hallucination occurred in a 12-category bar chart.}
\label{supp_VisualFaithfulness}
\resizebox{\textwidth}{!}{%
\begin{tabular}{|c|cccc|c|cccc|c|cccc|c|c|}
\hline
  & \multicolumn{5}{c|}{\textbf{Scatterplot}} 
  & \multicolumn{5}{c|}{\textbf{Line Chart}}
  & \multicolumn{5}{c|}{\textbf{Bar Chart}}
  &  \\
\cline{2-16}
Response Coding
  & SC & MC & SC-J & MC-J & All
  & SC & MC & SC-J & MC-J & All
  & SC & MC & SC-J & MC-J & All
  & All \\ 
\hline

Fully Faithful
  & 15 & 13 & 10 & 14 & 52
  & 14 & 15 & 15 & 11 & 55
  & 13 & 11 & 11 & 14 & 49
  & 156 \\

Partial Error
  & 0 & 2 & 5 & 1 & 8
  & 1 & 0 & 0 & 4 & 5
  & 1 & 4 & 4 & 1 & 10
  & 23 \\

Major Error
  & 0 & 0 & 0 & 0 & 0
  & 0 & 0 & 0 & 0 & 0
  & 1 & 0 & 0 & 0 & 1
  & 1 \\
\hline

\textbf{All}
  & 15 & 15 & 15 & 15 & 60
  & 15 & 15 & 15 & 15 & 60
  & 15 & 15 & 15 & 15 & 60
  & \textbf{180} \\
\hline
\end{tabular}
}
\end{table*}

Single-Class Juxtaposed (SC-J) charts frequently trigger the highest task counts, particularly in the unconstrained condition. 
Especially, in the unrestricted PC0 setting, all models utilize the greatest number of tasks.
However, in the most complex Multi-Class Juxtaposed (MC-J) configuration, we observe a notable decrease in task frequency compared to SC-J across all models (e.g., GPT decreases from 4.53 to 3.93; Claude from 5.26 to 4.60 in PC0). 
This counter-intuitive finding suggests that when faced with the combined density of multiple categories and panels, models transition from micro-level enumeration to macro-level synthesis. 
Rather than exhaustively decoding every interaction, the models prioritize global patterns that transcend individual panels, resulting in a more concise, higher-level interpretation despite the increased visual information.

\begin{table}[t]
\centering
\resizebox{\columnwidth}{!}{%
\begin{tabular}{c|c|ccc}
\toprule
Model & Chart Type &
PC2 $\rightarrow$ PC1 &
PC1 $\rightarrow$ PC0 &
PC2 $\rightarrow$ PC0 \\
\midrule

\multirow{4}{*}{Claude}
 & Bar chart      & 0.9256 & 0.8867 & 0.8658 \\
 & Line chart     & 0.9348 & 0.8722 & 0.8453 \\
 & Scatterplot    & 0.9079 & 0.8770 & 0.8625 \\
 & \textbf{Average} & \textbf{0.9228} & \textbf{0.8786} & \textbf{0.8579} \\
\midrule

\multirow{4}{*}{Gemini}
 & Bar chart      & 0.8982 & 0.8507 & 0.8116 \\
 & Line chart     & 0.8346 & 0.8559 & 0.7691 \\
 & Scatterplot    & 0.8570 & 0.8600 & 0.8137 \\
 & \textbf{Average} & \textbf{0.8633} & \textbf{0.8555} & \textbf{0.7981} \\
\midrule

\multirow{4}{*}{GPT}
 & Bar chart      & 0.8494 & 0.8830 & 0.8071 \\
 & Line chart     & 0.9081 & 0.8987 & 0.8643 \\
 & Scatterplot    & 0.8602 & 0.8978 & 0.8332 \\
 & \textbf{Average} & \textbf{0.8726} & \textbf{0.8932} & \textbf{0.8349} \\
\bottomrule
\end{tabular}
}
\caption{Semantic similarity scores across models and chart types, showing how consistently each LLM preserves core meaning under varying prompt constraints.}
\vspace{-15pt}
\end{table}

\begin{table}[t]
\centering
\resizebox{\columnwidth}{!}{%
\begin{tabular}{c|c|ccc}
\toprule
Model & Chart Type &
PC2 $\rightarrow$ PC1 &
PC1 $\rightarrow$ PC0 &
PC2 $\rightarrow$ PC0 \\
\midrule

\multirow{4}{*}{Claude}
 & Bar Chart       & 0.8358 & 0.8775 & 0.9250 \\
 & Scatterplot     & 0.8538 & 0.9795 & 0.9242 \\
 & Line Chart      & 0.9092 & 0.9283 & 0.9583 \\
 & \textbf{Average} & \textbf{0.8663} & \textbf{0.9284} & \textbf{0.9358} \\
\midrule

\multirow{4}{*}{GPT}
 & Bar Chart       & 0.9333 & 0.8900 & 0.9208 \\
 & Scatterplot     & 0.8598 & 0.9265 & 0.9432 \\
 & Line Chart      & 0.9042 & 0.8258 & 0.8417 \\
 & \textbf{Average} & \textbf{0.8991} & \textbf{0.8808} & \textbf{0.9019} \\
\midrule

\multirow{4}{*}{Gemini}
 & Bar Chart       & 0.9375 & 0.9458 & 0.9833 \\
 & Scatterplot     & 0.8571 & 0.9470 & 0.9886 \\
 & Line Chart      & 0.8657 & 0.9567 & 0.9567 \\
 & \textbf{Average} & \textbf{0.8868} & \textbf{0.9498} & \textbf{0.9762} \\
\bottomrule
\end{tabular}
}
\caption{
Consistency of statistical task retention as models expand from concise (PC2) to detailed (PC1, PC0) responses. All models maintain high retention across transitions and chart types, with Gemini showing the strongest consistency overall.}
\vspace{-12pt}
\label{Table/taskconsistancy}
\end{table}

\sloppy
\section{\textbf{Additional Qualitative Examples of Human vs LLM Reasoning Patterns}}
To supplement the analysis presented in the main text, we include additional examples that further illustrate the qualitative differences between how humans and LLMs interpret statistical graphics. 
Regarding a Google Stock chart (Figure~\ref{fig_major_error}(b)), one participant describes the visualization as a sequence of events: "The graph seems to be going continually upward... There is a spike right before 2008 and the fairly steep decline... then the increase continues again." 
Humans also tend to contextualize data by inferring causality based on external knowledge. 
For instance, in a chart showing programming proficiency, a participant hypothesizes that higher C\# proficiency among undergraduates \textit{"could be due to the languages being used during college"}, a context not explicitly present in the visualization.
These examples illustrate how human interpretation frequently extends beyond what is visually encoded, blending the observed pattern with prior knowledge to construct a coherent narrative.

For the same Google Stock chart, the LLM provides a static, hierarchical description: 
\textit{"The graph shows Google's stock price from October 2004 to January 2010, displaying a general upward trend from around \$100 to over \$700... followed by a significant decline to around \$300."} 
In another example regarding crop harvests, while a human roughly states, \textit{"Bakery products are the most harvested, fungi is the least"}, an LLM specifies, \textit{"Bakery Products having the highest harvest at approximately 25,000 tons, followed by Feedgrains at around 23,000 tons, while most other categories range between 8,000-20,000 tons."} 
Instead of using relative, qualitative terms like 'huge drop' or 'highest,' LLMs prioritize quantifying the visual space by citing specific values and ranges. 
Consequently, we interpret that humans perceive charts as a narrative (connecting points to tell a story), whereas LLMs perceive charts as a structure (measuring distances between elements and reading axis coordinates).
This interpretation reflects a systematic tendency to treat visual elements as measurable objects, focusing on numeric extraction.

% 문서 서두(preamble)에 반드시 아래 패키지를 추가해 주세요.
% \usepackage{tabularx}
\begin{table*}[t]
\centering
\footnotesize % 전체 폰트 크기를 약간 줄여서 공간 확보 (필요시 \scriptsize 로 변경)
\setlength{\tabcolsep}{3pt} % 열 사이의 기본 여백을 줄임
\caption{Codebook example using a spreadsheet to code all users' responses.}
\label{tab:codebook_example}
\begin{tabularx}{\textwidth}{|c|c|c|c|c|X|c|c|c|c|c|}
\hline
PID & Name & Graph & Data & Composition & Response-Description & Comprehension & Bloom's Taxonomy & Statistics & Design & Data \\ \hline
X & Horsepower & Scatterplot & Single class & Juxtaposed & 
Three different graphs to show the horsepower with miles per gallon to three manufacturing countries. & 
Complete Match & 
Comprehension & 
Correlation & 
NA & NA \\ \hline
\end{tabularx}
\end{table*}
\begin{table*}[t]
\centering
\resizebox{\textwidth}{!}{%
\begin{tabular}{c c | c c | c c | c c}
\toprule
\textbf{Human Task} & \textbf{Count} &
\textbf{LLM(PC0) Task} & \textbf{Count} &
\textbf{LLM(PC1) Task} & \textbf{Count} &
\textbf{LLM(PC2) Task} & \textbf{Count} \\
\midrule

Comparison & 61 &
Comparison & 165 &
Comparison & 149 &
Comparison & 118 \\

Trend & 48 &
Determine Range & 131 &
Trend & 104 &
Trend & 101 \\

Correlation & 47 &
Extremum & 131 &
Determine Range & 84 &
Determine Range & 75 \\

Compute Derived Value & 32 &
Trend & 120 &
Extremum & 84 &
Extremum & 73 \\

Extremum & 29 &
Characterize Distribution & 98 &
Correlation & 63 &
Correlation & 51 \\

Anomaly & 10 &
Correlation & 85 &
Characterize Distribution & 45 &
Characterize Distribution & 24 \\

Characterize Distribution & 8 &
Cluster & 67 &
Cluster & 24 &
Cluster & 15 \\

Cluster & 4 &
Anomaly & 55 &
Anomaly & 21 &
Anomaly & 12 \\

Determine Range & 4 &
Compute Derived Value & 18 &
Compute Derived Value & 11 &
Compute Derived Value & 4 \\

\bottomrule
\end{tabular}%
}
\caption{Ranking of statistical tasks by frequency: Human, PC0, PC1, and PC2.}
\label{tab:taskcount_rank_sorted}
\end{table*}

% 1. 톤다운된 커스텀 색상 정의
\definecolor{softcyan}{RGB}{116, 175, 185}   
\definecolor{softviolet}{RGB}{168, 162, 203} 
\definecolor{softorange}{RGB}{237, 182, 137} 

% 2. 컬러 매핑 매크로
\newcommand{\scolor}[2]{\cellcolor{softcyan!#1}#2}
\newcommand{\lcolor}[2]{\cellcolor{softviolet!#1}#2}
\newcommand{\bcolor}[2]{\cellcolor{softorange!#1}#2}
\newcommand{\gcolor}[2]{\cellcolor{gray!#1}#2}

\begin{table*}[t]
\centering
\renewcommand{\arraystretch}{1.0} 
\resizebox{\textwidth}{!}{%
\begin{tabular}{c|c|
cccc|c|cccc|c|cccc|c|c|}
\hline
% 헤더 수정: 첫 번째, 두 번째 컬럼을 합쳐서 Response Coding 표시
\multicolumn{2}{c|}{\multirow{2}{*}{\textbf{Response Coding}}} & 
\multicolumn{5}{c|}{Scatterplot} & 
\multicolumn{5}{c|}{Line Chart} & 
\multicolumn{5}{c|}{Bar Chart} & 
\multirow{2}{*}{All} \\
\cline{3-17}
\multicolumn{2}{c|}{} & 
SC & MC & SC-J & MC-J & All & 
SC & MC & SC-J & MC-J & All & 
SC & MC & SC-J & MC-J & All & 
\\ \hline

% ==========================================
% LLM SECTION
% ==========================================
\multirow{5}{*}{\rotatebox{90}{\textbf{LLM}}} 
 & Complete Match
 & \scolor{87}{13} & \scolor{73}{11} & \scolor{67}{10} & \scolor{53}{8} & \gcolor{15}{42}
 & \lcolor{87}{13} & \lcolor{60}{9} & \lcolor{87}{13} & \lcolor{53}{8} & \gcolor{15}{43}
 & \bcolor{73}{11} & \bcolor{73}{11} & \bcolor{60}{9} & \bcolor{80}{12} & \gcolor{15}{43}
 & \gcolor{20}{128} \\

 & General Match
 & \scolor{7}{1} & \scolor{20}{3} & \scolor{27}{4} & \scolor{27}{4} & \gcolor{15}{12}
 & \lcolor{13}{2} & \lcolor{7}{1} & \lcolor{7}{1} & \lcolor{27}{4} & \gcolor{15}{8}
 & \bcolor{20}{3} & \bcolor{27}{4} & \bcolor{27}{4} & \bcolor{20}{3} & \gcolor{15}{14}
 & \gcolor{20}{34} \\

 & Partial Match
 & \scolor{7}{1} & \scolor{7}{1} & \scolor{7}{1} & \scolor{20}{3} & \gcolor{15}{6}
 & \lcolor{0}{0} & \lcolor{33}{5} & \lcolor{7}{1} & \lcolor{20}{3} & \gcolor{15}{9}
 & \bcolor{7}{1} & \bcolor{0}{0} & \bcolor{13}{2} & \bcolor{0}{0} & \gcolor{15}{3}
 & \gcolor{20}{18} \\

 & No Match
 & \scolor{0}{0} & \scolor{0}{0} & \scolor{0}{0} & \scolor{0}{0} & \gcolor{15}{0}
 & \lcolor{0}{0} & \lcolor{0}{0} & \lcolor{0}{0} & \lcolor{0}{0} & \gcolor{15}{0}
 & \bcolor{0}{0} & \bcolor{0}{0} & \bcolor{0}{0} & \bcolor{0}{0} & \gcolor{15}{0}
 & \gcolor{20}{0} \\

\cline{2-18}
 & \textbf{Total}
 & \gcolor{10}{15} & \gcolor{10}{15} & \gcolor{10}{15} & \gcolor{10}{15} & \gcolor{25}{60}
 & \gcolor{10}{15} & \gcolor{10}{15} & \gcolor{10}{15} & \gcolor{10}{15} & \gcolor{25}{60}
 & \gcolor{10}{15} & \gcolor{10}{15} & \gcolor{10}{15} & \gcolor{10}{15} & \gcolor{25}{60}
 & \gcolor{35}{\textbf{180}} \\
\hline \hline

% ==========================================
% HUMAN SECTION
% ==========================================
\multirow{5}{*}{\rotatebox{90}{\textbf{Human}}} 
 & Complete Match
 & \scolor{75}{12} & \scolor{70}{11} & \scolor{50}{8} & \scolor{50}{8} & \gcolor{15}{39}
 & \lcolor{75}{12} & \lcolor{50}{8} & \lcolor{70}{11} & \lcolor{55}{9} & \gcolor{15}{40}
 & \bcolor{70}{11} & \bcolor{50}{8} & \bcolor{75}{12} & \bcolor{45}{7} & \gcolor{15}{38}
 & \gcolor{20}{117} \\

 & General Match
 & \scolor{20}{3} & \scolor{20}{3} & \scolor{10}{1} & \scolor{15}{2} & \gcolor{15}{9}
 & \lcolor{20}{3} & \lcolor{10}{0} & \lcolor{10}{1} & \lcolor{10}{1} & \gcolor{15}{5}
 & \bcolor{20}{3} & \bcolor{30}{5} & \bcolor{10}{1} & \bcolor{20}{3} & \gcolor{15}{12}
 & \gcolor{20}{26} \\

 & Partial Match
 & \scolor{25}{4} & \scolor{35}{6} & \scolor{50}{8} & \scolor{30}{5} & \gcolor{15}{23}
 & \lcolor{25}{4} & \lcolor{45}{7} & \lcolor{20}{3} & \lcolor{50}{8} & \gcolor{15}{22}
 & \bcolor{25}{4} & \bcolor{45}{7} & \bcolor{35}{6} & \bcolor{30}{5} & \gcolor{15}{22}
 & \gcolor{20}{67} \\

 & No Match
 & \scolor{30}{5} & \scolor{25}{4} & \scolor{45}{7} & \scolor{55}{9} & \gcolor{15}{25}
 & \lcolor{30}{5} & \lcolor{55}{9} & \lcolor{55}{9} & \lcolor{35}{6} & \gcolor{15}{29}
 & \bcolor{35}{6} & \bcolor{25}{4} & \bcolor{30}{5} & \bcolor{55}{9} & \gcolor{15}{24}
 & \gcolor{20}{78} \\

% 구분선 수정: 첫 번째 컬럼(Human)은 건너뛰고 2번째부터 줄 긋기
\cline{2-18}
 & \textbf{Total}
 & \gcolor{10}{24} & \gcolor{10}{24} & \gcolor{10}{24} & \gcolor{10}{24} & \gcolor{25}{96}
 & \gcolor{10}{24} & \gcolor{10}{24} & \gcolor{10}{24} & \gcolor{10}{24} & \gcolor{25}{96}
 & \gcolor{10}{24} & \gcolor{10}{24} & \gcolor{10}{24} & \gcolor{10}{24} & \gcolor{25}{96}
 & \gcolor{35}{\textbf{288}} \\
\hline
\end{tabular}
}
\caption{Comparison of comprehension match with designer's intent per chart type and composition (SC: single-class; MC: multi-class; SC-J: single-class juxtaposed; MC-J:multi-class juxtaposed). Top table: response alignment coding for LLMs, Bottom table: response alignment coding for Human. The match is classified into four categories: Complete Match, General Match, Partial Match, and No Match.}
\label{tab:comprehension_match}
\end{table*}

\end{document}